\documentclass[a4paper,fleqn,usenatbib]{mnras}

\usepackage{newtxtext,newtxmath}

\usepackage[dvipsnames]{xcolor}

\usepackage[T1]{fontenc}
\usepackage{ae,aecompl}

\usepackage{graphicx}	
\usepackage{amsmath}	
\usepackage{amssymb}	
\usepackage[flushleft]{threeparttable}
\graphicspath{{figs/}}
\usepackage{ulem}  

\usepackage{color} 

\usepackage{rotating} 
\usepackage{caption}
\usepackage{multirow, subfigure}
\usepackage{booktabs}
\usepackage{scalerel}
\usepackage{makecell}
\usepackage{verbatim}

\newcommand\ie    {{\it i.e.}}

\newcommand{\env}{\rm env}
\newcommand{\Rd}{R_{\rm d}}
\newcommand{\disc}{\rm disc}

\newcommand\eeq{\end{equation}}

\def\apj{ApJ}
\def\apjl{ApJL}
\def\aap{A\& A}
\def\nat{Nature}
\def\araa{ARA\&A}
\def\mnras{MNRAS}

\def\apjs{ApJS}

\def\mnras{{MNRAS}}

\def\pasp{{PASP}}

\definecolor{mex_pink}{RGB}{240,0,135}
\definecolor{emerald}{RGB}{0,145,80}

\title[3-D Radiative Transfer Modelling of W33A MM1]{Radiative transfer modelling of W33A MM1: 3-D structure and dynamics of a complex massive star forming region}
\author[Izquierdo et al.]{Andr\'es F. Izquierdo$^{1}$\thanks{E-mail: andres.izquierdo.c@gmail.com}, Roberto Galv\'an-Madrid$^{2}$\thanks{E-mail: r.galvan@irya.unam.mx}, 
Luke T. Maud$^{3}$, 
\newauthor
Melvin G. Hoare$^{4}$,
Katharine G. Johnston$^{4}$, 
Eric R. Keto$^{5}$ 
\newauthor
Qizhou Zhang$^{5}$ 
and
Willem-Jan de Wit$^{6}$ 
\\
\\
$^{1}$ Instituto de F\'isica - FCEN, Universidad de Antioquia, Calle 70 No. 52-21, Medell\'in, Colombia. \\
$^{2}$ Instituto de Radioastronom\'ia y Astrof\'isica,
            Universidad Nacional Aut\'onoma de M\'exico,
            Apdo. Postal 72-3 (Xangari), Morelia,\\
            Michoac\'an 58089, M\'exico. \\
$^{3}$ Leiden Observatory, Leiden University, PO Box 9513, NL-2300 RA Leiden, the Netherlands.  \\   
$^{4}$ School of Physics and Astronomy, University of Leeds, Leeds LS2 9JT, UK. \\
$^{5}$ Harvard-Smithsonian Center for Astrophysics, 160 Garden St, Cambridge, MA 02420, USA. \\
$^{6}$ European Southern Observatory, Alonso de C\'ordova 3107, Vitacura, Casilla, 19001, Santiago de Chile, Chile. \\
}

\date{Accepted 2018 April 20. Received 2018 April 20; in original form 2017 December 11
}

\pubyear{2018}

\begin{document}
\label{firstpage}
\pagerange{\pageref{firstpage}--\pageref{lastpage}}
\maketitle

\begin{abstract}
We present a composite model and radiative transfer simulations of the massive star forming core W33A MM1. The model was tailored to reproduce the complex features observed with ALMA at $\approx 0.2$ arcsec resolution in CH$_3$CN and dust emission. 
The MM1 core is fragmented into six compact sources coexisting within $\sim 1000$ au. In our models, three of these compact sources are better represented as disc-envelope systems around a central (proto)star, two as envelopes with a central object, and one as a pure envelope. The model of the most prominent object (Main) contains the most massive (proto)star ($M_\star\approx7~M_\odot$) and disc+envelope ($M_\mathrm{gas}\approx0.4~M_\odot$), and is the most luminous ($L_\mathrm{Main} \sim 10^4~L_\odot$). The model discs are small (a few hundred au) for all sources. 
The composite model shows that the elongated spiral-like feature converging to the MM1 core can be convincingly interpreted as a filamentary accretion flow that feeds the rising stellar system. The kinematics of this filament is reproduced by a parabolic trajectory with focus at the center of mass of the region. Radial collapse and fragmentation within this filament, as well as smaller filamentary flows between pairs of sources are proposed to exist.
Our modelling supports an interpretation where what was once considered as a single massive star with a $\sim 10^3$ au disc and envelope, is instead a forming stellar association which appears to be virialized and to form several low-mass stars per high-mass object. 

\end{abstract}

\begin{keywords}
stars: formation -- stars: protostars -- stars: massive -- radiative transfer 

\end{keywords}



\section{Introduction} \label{sec:intro}
The formation of stars can occur in different environments, ranging from isolated to highly clustered systems \citep{LadaLada03}. There is evidence that the more massive the stellar system is, the less likely it is to form in isolation \citep{Sana17}. Therefore, improving our understanding of intermediate- and high-mass star formation comes together with our knowledge of the formation of multiple stellar systems. A recent review that emphasizes the link between the formation of massive stars and their clusters is presented in \cite{Motte17}. 

Earlier interferometric observations showed that massive stars form through accretion from structures that could be rotationally supported discs \citep[e.g.,][]{Cesaroni99,Zhang02,Patel05,CG12}. However, the advent of the Atacama Large Millimeter/submillimeter Array (ALMA) is changing the landscape of star formation research by providing unprecedented high angular resolution, sensitivity, and dynamic range images of the participating dust and gas. One of the overall conclusions that can be obtained from considering recent results is that {\it a few} intermediate and massive stars can form as scaled up versions of the low-mass star formation paradigm: a {\it single} Keplerian disc -- which could be circumbinary -- plus a rotating/infalling envelope at early stages \citep[e.g.,][]{SanchezMonge13,BeltranDeWit16,Girart18}; whereas {\it many} massive stars form in {\it clustered} systems at clump 
\citep[$\sim 0.1$ to 1 pc;][]{Liu15}
or even core 
\citep[$< 0.1$ pc;][]{Johnston15,Hunter17,Beuther17,Maud+17} 
scales. \cite{Cesaroni17} find evidence for Keplerian discs in about half of their small sample. In contrast, \cite{Ginsburg17} find no evidence of discs in a more highly clustered and luminous star formation region.  

Radiative transfer simulations are needed to interpret the complexity of current observations. A variety of public codes to calculate the (sub)mm line and continuum emerging from 3D models have been presented and tested in the literature, e.g., MOLLIE \citep{KetoRyb10}, and LIME \citep{Brinch+10}. Some of these codes provide basic model setups, but composite 3D models are often needed to better represent complex structures. In this spirit, efforts to produce radiative transfer models of star forming systems with multiple components have recently appeared in the literature \citep{Schmiedeke16,Quenard17}. 

In this paper, we present a multiple-component radiative transfer model that aims at reproducing the main features observed with ALMA in the high-mass star formation core W33A MM1 \citep[e.g.,][]{Maud+17,GM10}. 
This core is the most massive in W33A and hosts the most luminous Young Stellar Object (YSO), traced by a faint hypercompact HII region \citep{vdTMenten05}. Further evidence for at least one massive ($M > 10~M_\odot$) YSO in MM1 comes from high angular resolution IR observations \citep{Bunn95,deWit07,deWit10,Davies10}. 
Previous sub-arcsecond Submillimeter Array (SMA) observations pointed toward the existence of a massive gaseous disc of a few $M_\odot$ surrounding a potentially massive ($M_\star \sim 10~M_\odot$) YSO centered in the millimeter source Main within MM1 \citep{GM10}. 

W33A is part of the W33 molecular cloud complex \citep{vdT00,Immer14,Lin16}. Its parallax distance to the Sun has been measured to be $2.4^{+0.17}_{-0.15}$ kpc \citep{Immer13}. 

The ALMA observations that we model here were presented in \cite{Maud+17}. These data has $\times 3$ better angular resolution and $\times 15$ better sensitivity than our previous SMA observations. The ALMA images reveal that what we previously thought was a massive rotating disc, probably with one unresolved companion, is actually a multiple system in formation, although the kinematics is still dominated by the most massive object MM1 Main. A prominent spiral-like filamentary gas stream appears to feed the central part of MM1 from the northwest.

The outline of the paper is as follows: 
Section \ref{sec:observations} describes the observations that we model. Section \ref{sec:physmodel} explains the individual physical components that are used. Section \ref{sec:Grid} details on the construction of the composite 3D grid. Section \ref{sec:RT} describes the implementation in LIME. \ref{sec:model_building} describes the logical order in which the final global model was found. Section \ref{sec:Model_parameters} explains the determination of the model parameters. Sections \ref{sec:K4} and \ref{continuum} give the results of the line and continuum model, including a comparison to observations. Sections \ref{sec:discussion} and \ref{sec:conclusions} are a discussion of the results and the conclusions, respectively. Appendix \ref{ap:chanmaps} shows the channel map emission in the models and observations. Finally, Appendix \ref{ap:models} gives information on how to access the tools that we developed to create the complex models and how to use them within LIME, which we believe can be of interest to the community. 

\section{Observational Data} \label{sec:observations}

The ALMA observations were originally scheduled as an A-ranked Cycle 1 project (2012.1.00784.S -- PI: M. G. Hoare), but due to the need for the then longest baselines, they were not successfully executed until June 2015. For more details of the data set we refer to \cite{Maud+17}.

Due to the multiplicity in the region within a few arcseconds, we modelled only the data that are less morphologically confused, and lines that do not appear to be spectroscopically blended or contaminated by others. Therefore, we selected the CH$_3$CN $J=19-18$ $K=4$ and $K=8$ lines, as well as the 0.8 mm (Band 7) continuum. We also use the 1.3 mm (Band 6) continuum for further comparisons with our models, although it has a slightly lower angular resolution. 

The 1.36 mm (220.818 GHz) continuum image has a synthesized beam FWHM $=0.33\times0.24$ arcsec, with a position angle PA $=-46.2^\circ$. The rms noise in this image is $\sigma_\mathrm{rms,1.3mm} \approx 112~\mu$Jy beam$^{-1}$ (0.035 K). The 0.86 mm (349.454 GHz) continuum image has an FWHM $=0.21\times0.14$ arcsec, PA $=-80.9^\circ$, and $\sigma_\mathrm{rms,0.8mm} \approx 177~\mu$Jy beam$^{-1}$ (0.059 K). The $K=4$ ($\nu_0=349.3463$ GHz) and $K=8$ ($\nu_0=349.0249$ GHz) CH$_3$CN cubes have almost identical beams with FWHM $=0.20\times0.14$ arcsec and a PA $=-79.8^\circ$. Their noise levels are  $\sigma_\mathrm{rms,K4} \approx 3.7~$mJy beam$^{-1}$ (1.28 K) and $\sigma_\mathrm{rms,K8} \approx 2.1~$mJy beam$^{-1}$ (0.73 K), respectively, in channels 0.42 km s$^{-1}$ wide. 

\subsection{Main observational features} \label{sec:obsmotivation}

In this section we enumerate the main observational features that motivated the components of our MM1 model. Further motivation will become apparent through the rest of the paper. 

\begin{figure*}  
 \includegraphics[width=0.9\textwidth]{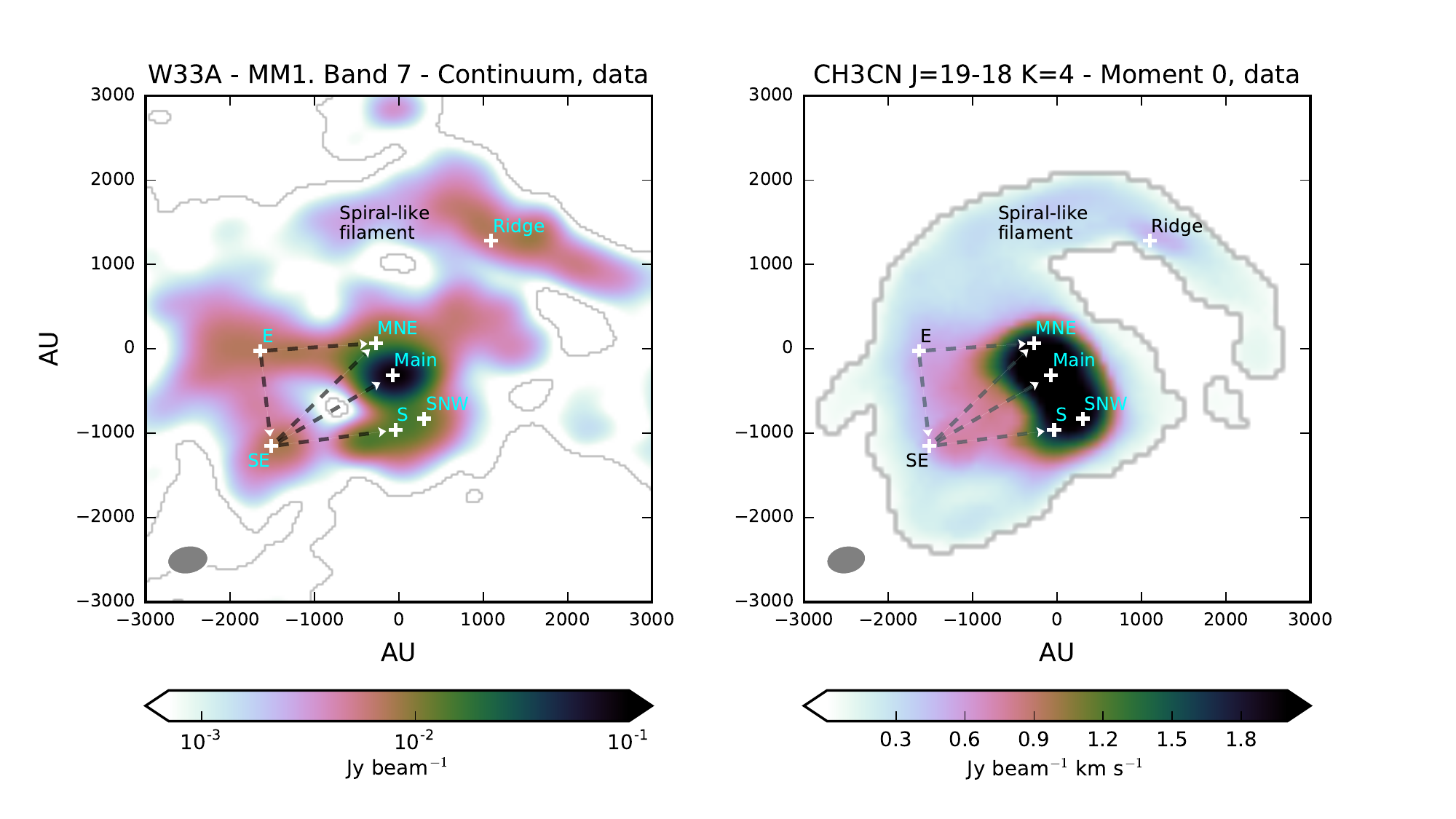} 
 \caption{ \textit{Left}: ALMA 349 GHz (0.8 mm) continuum image of W33A MM1. \textit{Right}: velocity-integrated intensity (moment 0) of the CH$_3$CN $J=19-18$ $K=4$ line data. Cells with intensity below 10\% of the peak were masked out. The compact sources included in the model are marked by crosses ($+$). The dashed arrows represent the modelled inter-source filaments and their flow direction. A label indicates the zone where the spiral-like filament resides. The beam size is shown in the lower left corner of the panels.} 
\label{fig:obs_features}
\end{figure*}

In Figure \ref{fig:obs_features} we present the Band 7 continuum emission and the CH$_3$CN $J=19-18$ $K=4$ moment 0 maps for the observational data. There, we highlight the compact sources as reported in \cite{Maud+17} and \cite{GM10}: Main, South (S), South-East (SE), East (E) and Ridge. Two new compact sources are proposed in this paper and are labeled under the same rules: Main North-East (MNE) and South North-West (SNW), as well as five inter-source filaments represented with dashed lines, and the spiral-like filament in the northern part of the region. 

The existence of MNE and SNW is mainly motivated by the CH$_3$CN line emission. The moment 0 map evidences extended emission toward the northeast of Main and the northwest of S. This emission is also warm \citep[see][]{Maud+17}, and its velocity field is more consistent with separate, redshifted blobs (see the last row of Fig. \ref{fig:channels}) than with a single, more extended source. For more details see Section \ref{sec:K4}.
On the other hand, the existence of inter-source filaments is motivated mostly by the extended features in the continuum maps that appear to join the compact sources (see Figure \ref{fig:obs_features} and Section \ref{sec:acc_filaments-results}), although some of these features are also apparent in the CH$_3$CN maps. 

\section{Model} \label{sec:model}

\subsection{Physical models} \label{sec:physmodel}

In order to describe this complex star formation region, we produce a global model made of the superposition of individual components, all within a 7000 au cubic region representing W33A MM1. In this section we describe the physical attributes of the individual components. We emphasize that the final model is not unique nor a best fit, but it provides a reasonable match to the data and is physically motivated. The justification for the use and characteristics of each element are further explained in Sections \ref{sec:model_building},  \ref{sec:Model_parameters}, and \ref{sec:results}.

\begin{figure}  
 \includegraphics[width=\columnwidth]{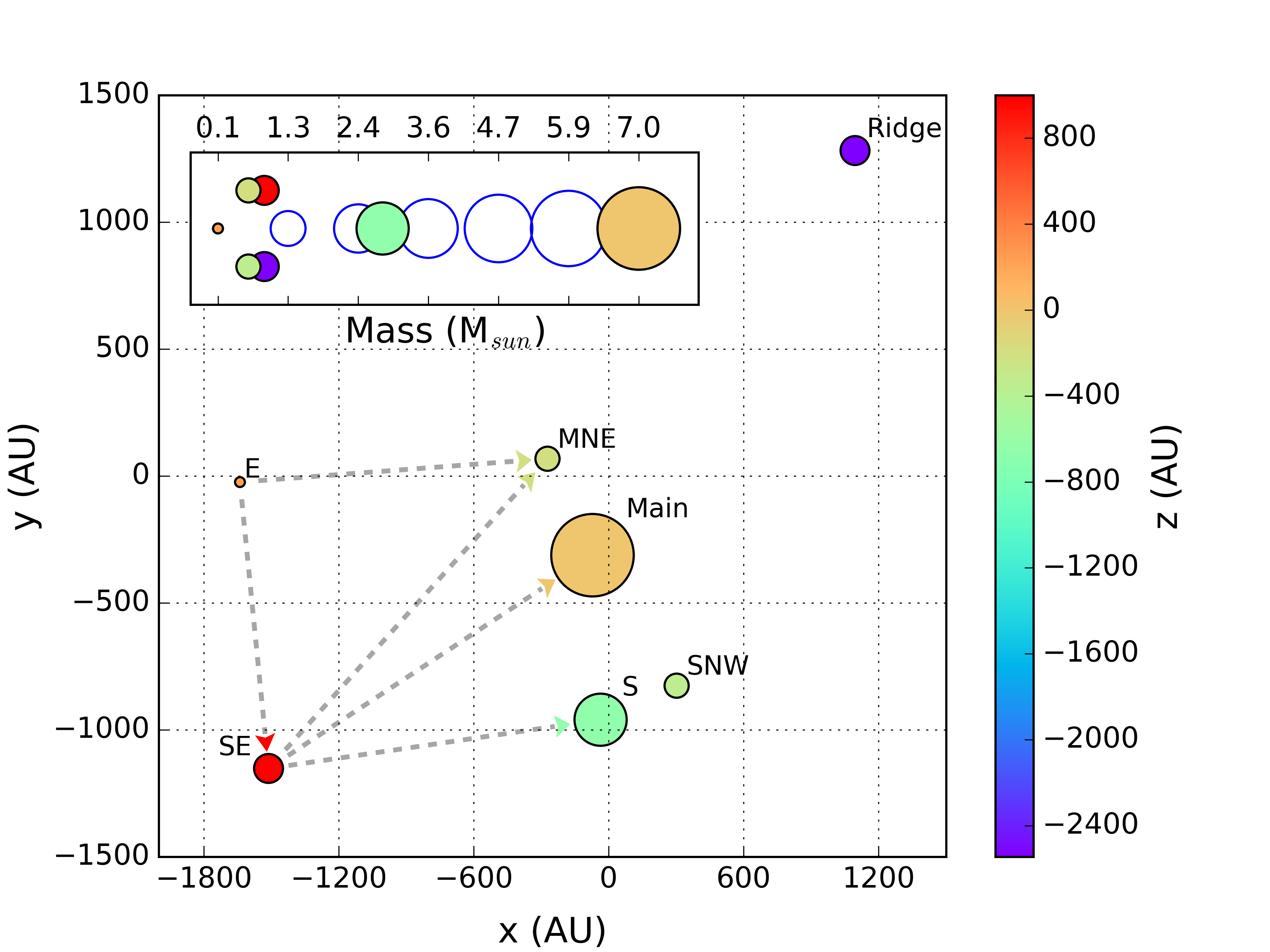} 
 \caption{Configuration of compact sources and inter-source filaments in the W33A-MM1 model. The $x$ and $y$ axes are parallel to $RA$ and $Dec$. The $z$-axis is parallel to the line of sight, with positive values increasing away from the observer. The z = 0 plane intersects the position of Main, and $(x,y)=(0,0)$ coincides with the phase center of the ALMA images. Arrows indicate the direction of the gas flows between sources. The sizes of the markers are related to the mass of the sources as shown in the top subplot.} 
\label{fig:sketch}
\end{figure}

From the seven compact sources, four of them are disc-envelope systems (Main, S, SE, Ridge), two of them are pure rotating/infalling envelopes (MNE, SNW), corresponding to less evolved YSOs, and one is a turbulent sphere (E), corresponding to an even younger object. Figure \ref{fig:sketch} shows a schematic diagram of the relative positions and central masses of these sources. 

Additionally, we include five cylindrical filaments joining pairs of compact sources (also sketched in Fig. \ref{fig:sketch}), plus a larger spiral-like filament which we interpret as an accretion flow feeding the center of MM1 from its periphery, as proposed by \cite{Maud+17}.

\subsubsection{Compact Sources} \label{sec:compactsources}

We implemented a standard YSO modelling for the four disc-envelope sources, following the approach of \cite{KetoZhang10}. Those authors superpose a rotationally supported flared disc embedded in an infalling and rotating envelope. The envelope is modelled using the prescription of \cite{Ulrich76}, who constructs the density and velocity fields assuming that the particles around the stellar source follow ballistic paths. Although simple, this model has been useful to reproduce observations from the scales of low-mass (proto)stars up to high-mass clusters 
\citep[e.g.,][]{Whitney03,Keto07,Maud13b}.

For the envelope density we use equation 1 of \cite{KetoZhang10}: 

\begin{equation} \label{eq:1}
\begin{split}
\rho_{\env} (r,\theta) = &\rho_{\rm e_0} (r / \Rd)^{-3/2} \left(1 + \frac{\cos \theta}{\cos \theta_0}\right) ^ {-1/2} \\ 
&\times [1+(r/\Rd)^{-1}(3\cos^2\theta_0-1)] ^ {-1},  
\end{split}
\end{equation}

\noindent
where $r$ is the distance to the center of the model and $\theta$ the polar angle; $\Rd$ is the centrifugal radius, defined as an equilibrium zone where the gravitational force of the central-point mass is equal to the centrifugal force of the rotating envelope;  $\theta_0$ is the initial angle of the streamline and $\rho_{\rm e_0}$ is the envelope normalization density at $r=\Rd$ and $\theta=\pi/2$. They must satisfy the geometrical relation 
\citep[equation 2 of][]{KetoZhang10}:

\begin{equation} \label{eq:2}
r = \frac{\Rd \cos\theta_0 \sin^2\theta_0}{\cos\theta_0 - \cos\theta}.\\
\end{equation}

This constraint can be used to find an analytical expression for $\cos\theta_0$ \citep[see its functional form in equation 13 of][]{Mendoza04}. 

The velocity components of the envelope are 
\citep[see equations 14--16 of][]{KetoZhang10}: 

\begin{align} 
&v_r(r,\theta) = - \left(\frac{G M_\star}{r} \right)^{1/2} \left(1+\frac{\cos\theta}{\cos\theta_0} \right)^{1/2}, \label{eq:3}\\
&v_\theta(r,\theta) = \left(\frac{G M_\star}{r} \right)^{1/2} \left(\frac{\cos\theta_0 - \cos\theta}{\sin\theta} \right)^{1/2} \left(1+\frac{\cos\theta}{\cos\theta_0} \right)^{1/2}, \label{eq:4}\\
&v_\phi(r,\theta) = \left(\frac{G M_\star}{r} \right)^{1/2} \frac{\sin\theta_0}{\sin_\theta}\left(1-\frac{\cos\theta}{\cos\theta_0} \right)^{1/2} \label{eq:5}
\end{align}

For the temperature of the envelope we follow \cite{Whitney03} and assume $T_{\env}(r)\propto r^{-0.33}$. 
The proportionality constant is left as a free parameter to adjust density weighted mean temperatures according to the observational data.  

We include the possibility of adding a conical cavity with an arbitrary opening angle. \\

\begin{figure*}
\centering
\subfigure[MNE. \label{fig:MNE}]
{\includegraphics[scale=0.6]{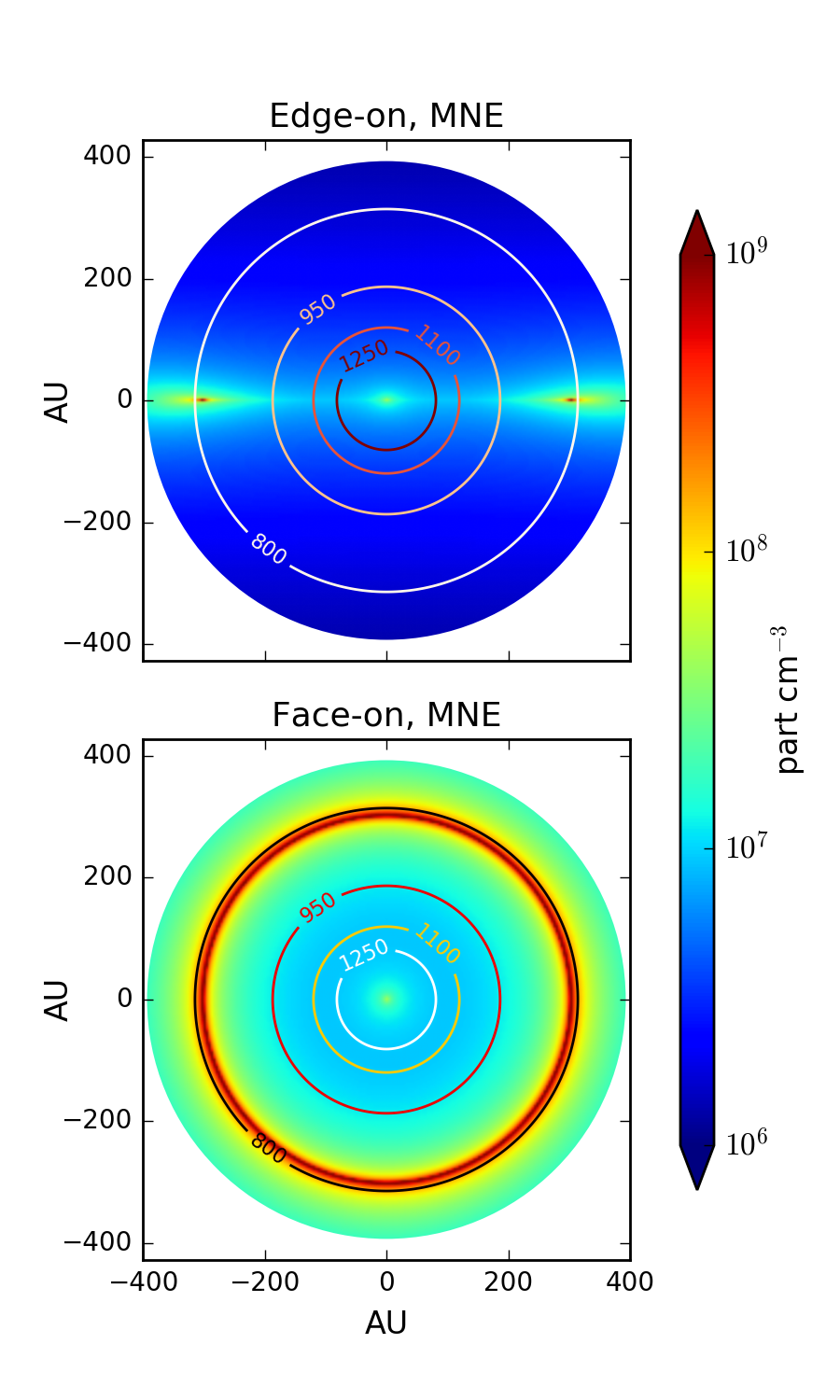}} 
\subfigure[Main.  \label{fig:Main}]{\includegraphics[scale=0.6]{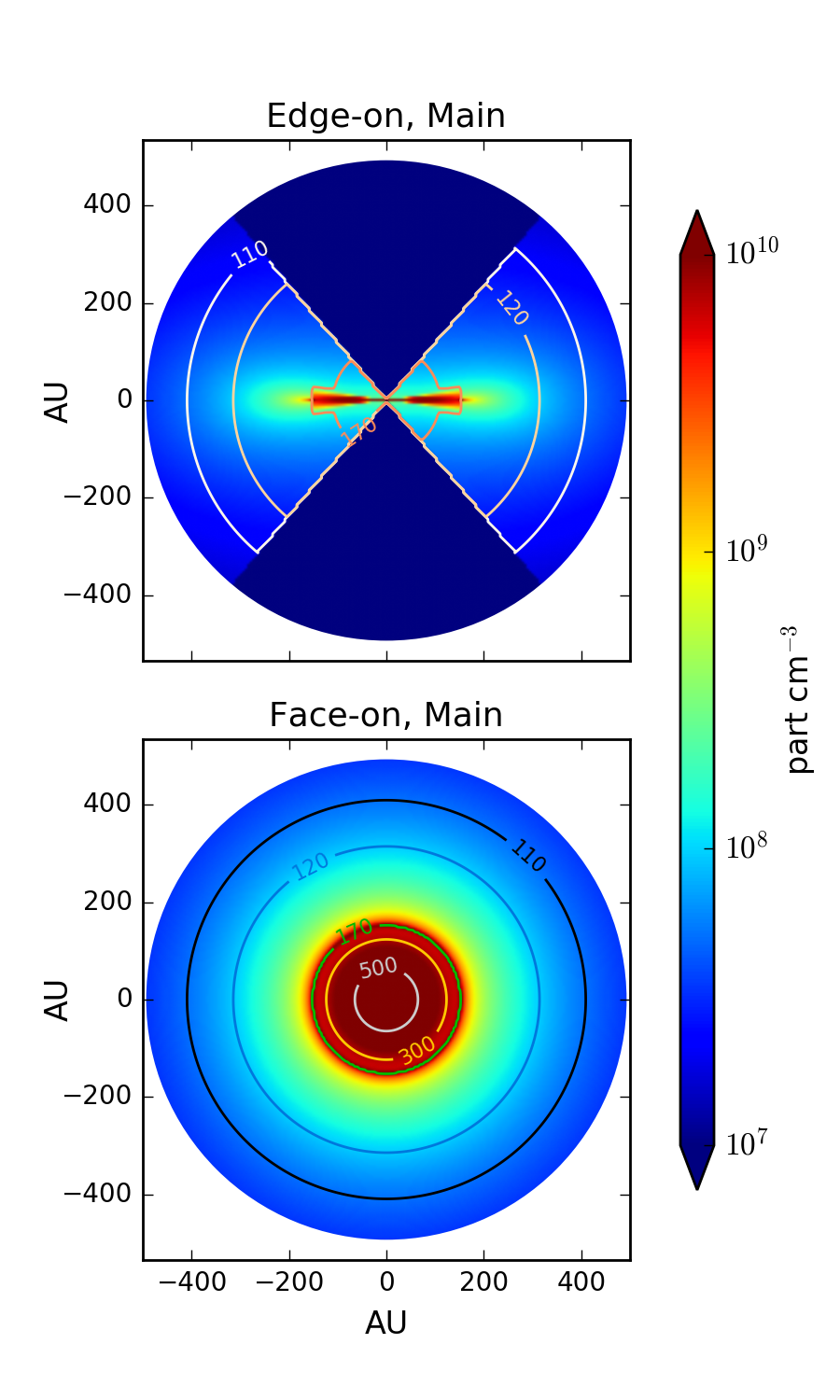}} 
\caption{{\it (a) Left:} density (colors) and temperature (contours) profiles for a pure (Ulrich) envelope compact source. The example corresponds to the final model for MNE. {\it (b) Right:} density and temperature profiles for a compact source made of a disc plus an envelope. The example corresponds to Main, which has a cavity. The top subplots show the edge-on ($i=90^\circ$) middle cut of the models, and the bottom subplots show the face-on ($i=0^\circ$) middle cut.} \label{fig:disc_envelope}
\end{figure*}

For the disc, we implemented the standard prescription by \cite{Pringle81}. We assume a steady, Keplerian, flared disc limited within the centrifugal radius $\Rd$ of the envelope. The disc density field is given by equation 3.14 of \cite{Pringle81}: 

\begin{subequations}\label{eq:6}
\begin{equation} \tag{6}
\rho_{\disc} (z,R) = \rho_0 (R) \exp\{-z^2 / 2H^2(R)\}, 
\end{equation} \\
where $R$ and $z$ are cylindrical coordinates. $H(R)$ is the scale-height of the disc and $\rho_0(R)$ is the disc density function in the mid-plane \citep[equations 7 and 8 of][]{KetoZhang10}: 

\begin{gather} 
H(R) = H_{0} (R / R_\star)^{1.25}, \label{eq:6a}\\
\rho_{0} (R) = A_\rho\rho_{\rm e_0}(\Rd/R)^{2.25}; \label{eq:6b} 
\end{gather}
\end{subequations}
\\
\noindent
The scale-height at the stellar radius is set to $H_{0} = 0.01 R_\star$, and $A_\rho$ is the density factor between disc and envelope at $\Rd$. The disc velocity is 
\citep[equation 3.3 of][]{Pringle81}: 

\begin{equation} \label{eq:7}
\vec{v}_{disc} =  \sqrt{GM_\star / R} \hspace{2mm} \hat{\phi},\\
\end{equation}
\\
and the temperature \citep[equation 12][]{KetoZhang10}: 

\begin{equation} \label{eq:8}
T_{\disc} = B_{\rm T} \left[ \left( \frac{3GM_\star \dot{M}}{4\pi R^3\sigma} \right) \left( 1- \sqrt{\frac{R_\star}{R}}\right) \right] ^{1/4},
\end{equation}
\\
where $B_{\rm T}$ is a factor to adjust disc heating and $\dot{M}$ is the mass accretion rate given by equation 3 of \cite{KetoZhang10}: 
\begin{equation} \label{eq:9}
\dot{M} = \rho_{\rm e_0} 4\pi \Rd^2  v_{\rm k}, 
\end{equation}
\noindent
where $v_{\rm k}$ is the Keplerian velocity at $\Rd$.\\

Figure \ref{fig:disc_envelope} shows the density and temperature structure of example disc and envelope models. 

\subsubsection{Accretion filaments} \label{sec:filaments}

Elongated features joining some pairs of compact sources are apparent in the continuum images and in the line cubes (see Section \ref{sec:results}). We modelled them as `accretion filaments'. Most of them are implemented as straight cylinders of $\sim 10^3$ au length that join pairs of compact sources, and with uniform density and temperature. The accompanying code leaves open the option of adding a dependency of density and temperature with cylindrical radius. We assume the kinetic energy of the filaments is related to their gravitational potential energy as given by the virial theorem:

\begin{equation} \label{eq:10}
\frac{3}{5}\frac{GM_\star}{r} = \frac{1}{2} v^2,
\end{equation}
\\
from which we obtain the speed at each point of the model. The main axis of each cylinder is defined as  $\vec{r}_{\rm cyl-ax} = \vec{r}_{\star^>} - \vec{r}_{\star^<}$, where $\vec{r}_{\star^>}$ and $\vec{r}_{\star^<}$ are the positions of the most and least massive compact object in each pair. Additionally, we consider that the gravitational potential is fully determined by the most massive of the pair of sources and that the velocity in the cylinder points towards that source. This is analogous to considering that the entire gas inside the filament is within the Hill radius of the most massive compact source. The velocity field for each cylinder can be written as: 

\begin{equation} \label{eq:11}
\vec{v}_{\rm cyl}(\vec{r}) = \left( {\frac{6 GM_{\star^>}}{5} } \right)^{1/2}  \frac{\vec{r} - \vec{r}_{\star^>}}{|\vec{r} - \vec{r}_{\star^>}|^{3/2}}. 
\end{equation}
\\
\noindent
Since our simplifying assumptions imply that the most massive compact source is taking material from the least massive one, we 
fixed the systemic (initial) velocity of the flows to be the same as that of the low mass source in each pair. Note that the previous assumptions automatically fix the relative orientation of compact sources in the z (line of sight) axis. 

There are 5 filamentary flows between pairs of compact sources in our model: SE$\rightarrow$S, SE$\rightarrow$Main, SE$\rightarrow$MNE, E$\rightarrow$MNE and E$\rightarrow$SE (see Fig. \ref{fig:sketch}). 

Additionally to the small cylindrical filaments, we include a larger (7240 au length) accretion filament reaching the central part of MM1 from its north side, following the interpretation of \cite{Maud+17} that this spiral-like structure is a `feeding filament' that deposits material to the central region of MM1. 
We model the feeding filament structure using a parabolic cylinder with focus at the center of mass of the entire region. 

The speed along the parabolic trajectory is given by orbital energy conservation, and we also include 
a velocity component of collapse toward the axis of the parabolic filament. Thus, the vector velocity is 

\begin{equation} \label{eq:12}
\vec{v}_{\rm par}(\vec{r}) = \left( {\frac{2 GM_{\rm c}}{|\vec{r} - \vec{r}_{\rm cm}|} } \right)^{1/2} \hat{t} + v_{\rm in} \hat{n},  
\end{equation}

\noindent
where $\hat{t}$ and $\hat{n}$ are the tangent and normal unitary vectors associated to the main axis of the parabola. We set the infall velocity ($v_{\rm in}$) within the parabolic filament in terms of the speed of sound in the medium: $c_{s} = (\gamma k_B T/2m_{\rm H})^{1/2}$, where $\gamma$ is the characteristic heat capacity ratio of the medium, $k_B$ the Boltzmann constant and $m_{\rm H}$ the Hydrogen mass. More details can be found in Section \ref{sec:acc_filaments-results}.   

For simplicity, we set the density and temperature of the filaments to be homogeneous in each of them.

\begin{figure}  
 \includegraphics[width=\columnwidth]{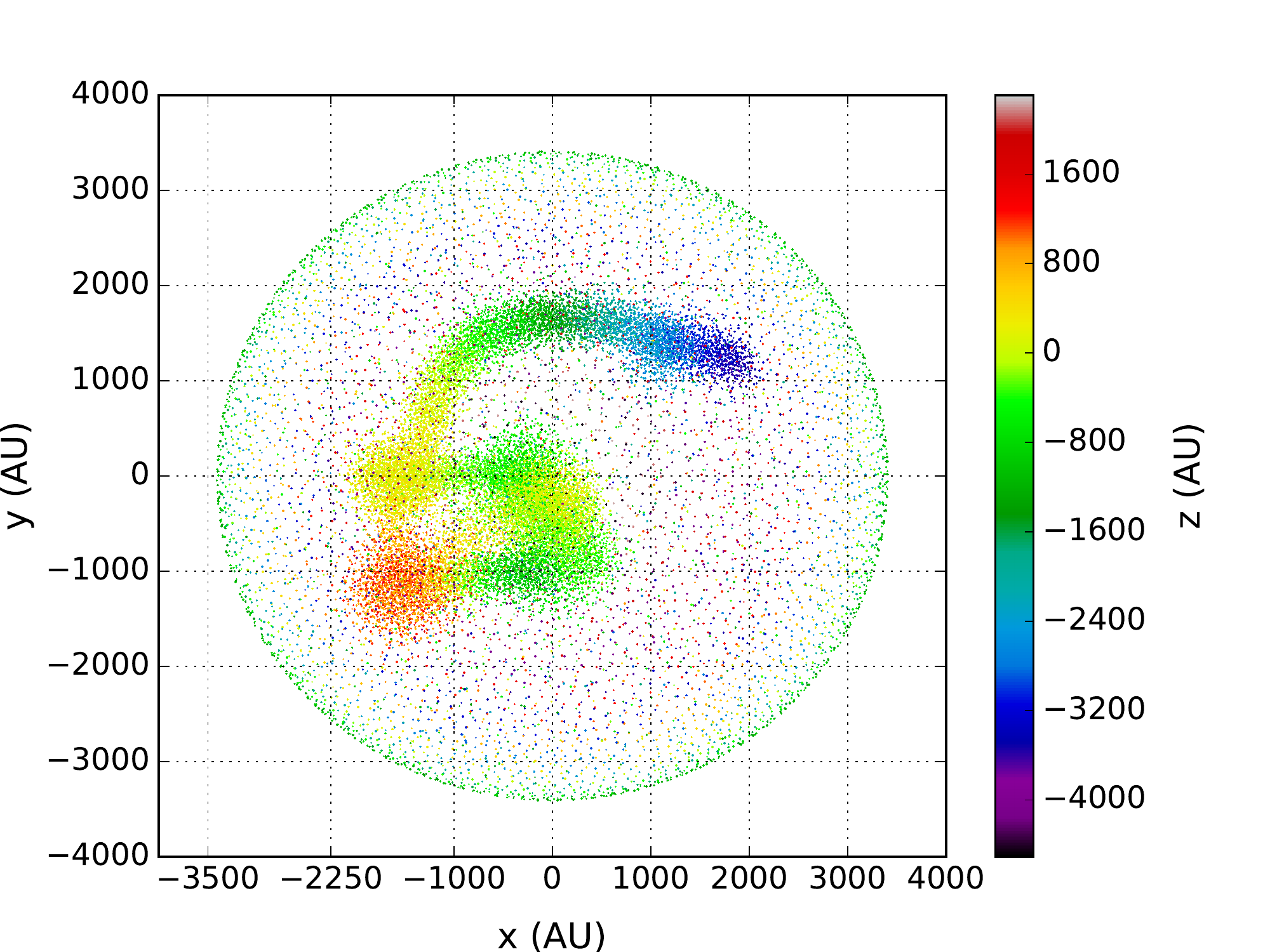} 
 \caption{Visualization of the global grid, which is made of the superposition of many local grids, each representing an individual sub-model. Regions with a higher density of points have a larger mass density in the model. Colors indicate the depth in the $z$ axis, with  positive values increasing away from the observer. The z = 0 plane intersects the position of Main, and $(x,y)=(0,0)$ coincides with the phase center of the ALMA images.}
\label{fig:GridPoints}
\end{figure}

\subsection{The model grids} \label{sec:Grid}

\subsubsection{Global grid} \label{sec:GlobalGrid}

We create a \textit{global grid} that harbors the individual \textit{local grids} (models) together. 
Each local grid is rotated and translated within the global grid according to the  restrictions given by observational data. The global grid is homogeneous and cubic: it has 301 nodes distributed in equal steps over 7000 au in each direction, \ie, it is built with $\sim 27 \times 10^6$ points, and its linear resolution is 23.3 au.  Figure \ref{fig:GridPoints} shows a visualization of the final global grid where the density of plotted points is proportional to the model density. 

To overlap the \textit{local grids} within the global grid, an algorithm for distance minimization between the nodes of both was made. Thus, spatial information is extracted from each node of each local grid and the corresponding nearest node in the global grid is found. This point will inherit the physical properties that the local grid point was saving. 
Since local grids are generally more spatially resolved than the global grid, it is likely to occur that some cells of {\it a given} local grid converge to the same closest node in the global grid. When this happens, the latter node takes the average of the overlapping densities. Other properties are taken as their density-weighted averages. 

After the previous step is done, the algorithm similarly checks whether two or more nodes of \textit{different} local models collapse into the  same node in the global grid. This time, the latter node takes the sum of the densities (for mass conservation) and other properties are again  density-weighted averages.  

\subsubsection{Local grids} \label{sec:LocalGrids}

For the compact sources, each local grid was also set to be homogeneous in Cartesian coordinates. These local grids can have different physical sizes and resolutions each. 
 
For the models containing an Ulrich envelope, we added a condition to ensure that no point of their grids falls into the mathematically undefined position $(r, \theta) = (\Rd, \pi/2)$, where the density diverges. The interpretation of this jump in density is that the `true' disc starts inwards \citep{Mendoza04}. 

For the cylinders, the nodes of their local grids are evenly and randomly distributed. To do so, we first generate a random point along the axis of the cylinder. Secondly, we create a vector with fixed position in that point, with random orientation (restricted to be perpendicular to the axis) and length in the ranges $[0,2\pi]$ and $[0, R_{\rm cyl}]$, respectively. This recipe is repeated $(2R_{\rm cyl})^2 |\vec{r}_{\rm cyl-ax}| / dr^3$ times to ensure that all the cells of the global grid enclosed by the virtual cylindrical surface will have enough points around them to be filled with. $dr$ is the maximum possible separation between neighboring nodes in the global grid: $dr = \sqrt{dx^2 + dy^2 + dz^2}$. 

The grid of the parabolic filament was built in a similar way to those of the cylinders, with an extra consideration due to the curved trajectory. First, we consider the characteristic equation of a parabola is $x^2 = 4py$, where $p$ is the parameter that accounts for its focus. Then, it is possible to calculate all the tangent vectors in the parabolic section of analysis as follows:

\begin{equation} \label{eq:14}
\begin{aligned}
&\hat{t} = \cos(\theta)\hat{i} + \sin(\theta)\hat{j}, \\
&\theta(x) = \tan^{-1}(x / 2p), 
\end{aligned}
\end{equation}
if its vertex is located in (0,0). Each of these tangent vectors represent a local main axis around which we generated a random point as in the second step of the construction of cylindrical grids. We repeat this step enough times to populate correctly this zone in the global grid, as explained in the previous paragraph. 

\subsection{Radiative transfer simulations} \label{sec:RT}

We use version 1.6.1 of the Line Modelling Engine code \citep[LIME,][]{Brinch+10} to perform radiative transfer simulations of the physical grids described in the previous section. LIME calculates both molecular line and (dust) continuum maps by solving the molecular excitation or fixing it in the LTE case, and then the transfer of radiation through the model. It builds unstructured 3D Delaunay grids by generating a set of random points across the domain that will be accepted or rejected according to the density of the given model. Then, the grid is smoothed via Voronoi tessellations. LIME retrieves molecular data from the LAMDA database \citep{Schoier+05}.

A few adaptations had to be made to be able to feed our models into LIME. The necessary code is available through the GitHub link provided in Appendix \ref{ap:models}. We included a header script in LIME to upload the output data from our model-generating codes. Also, we added in the user interface script (\textit{model.c}) an algorithm to determine the nearest neighbors between the randomly generated set of points in LIME and the points of our global grid, following the suggestions made in the LIME documentation\footnote{\href{http://lime.readthedocs.io/en/latest/}{http://lime.readthedocs.io/en/latest}, section \textit{Advanced setup.}}. Given that our grid is homogeneous in Cartesian coordinates, it is possible to efficiently determine the closest pairs of points. Let us call a random generated LIME point $(x_r,y_r,z_r)$ and its nearest point in our grid $(x_m,y_m,z_m)$. First, we compute the nearest $yz$ plane associated with the given $x_r$, so, $x_m$ is found. Then, in that plane we look for the nearest column associated with the given $y_r$, so $y_m$ is found. Finally, in that column we compute the nearest cell associated with the given $z_r$, so $z_m$ is found. In the end, the point $(x_r,y_r,z_r)$ receives all the properties belonging to $(x_m,y_m,z_m)$. \\

The rotational $(J,K)$ transitions of CH$_3$CN are such that several $K$ lines for a given $J+1 \rightarrow J$ can be observed in a single spectral setup \citep[e.g.,][]{Cummins83,Remijan04}. This fact has made these lines a widely-used tracer of warm, dense gas \citep[e.g.,][]{Araya05,Purcell05,Cesaroni17}. We use LTE calculations for our modelling. This is justified since the critical density of the $J=19-18$, $K=4$ and $K=8$ transitions at the model temperatures is $n_{\rm crit} \approx 1\times 10^7$ cm$^{-3}$. Also, `effective' densities for thermalization are typically at least one order of magnitude below critical densities \citep{Evans99}. Most of the mass in our domain is above the critical density of the modelled lines. 

We allow for different CH$_3$CN abundances with respect to H$_2$ for each sub-model (see Section \ref{sec:Model_parameters}). The gas-to-dust mass ratio was set to 100 in the entire global grid.

Some fluctuations in the model continuum emission appear because the grid points randomly generated by LIME\footnote{Defined with the `pIntensity' parameter.} do not map the region completely, since they are fewer than the model grid points. Therefore, different model regions are better covered with LIME grid points in some runs than in others. To smooth these fluctuations, for each continuum image presented in this paper, we generated ten images with the same set of parameters.  Then we extracted the median of the intensity for each pixel and constructed a final image. This averaging process is equivalent to generate an image with a higher number of grid points in LIME, but faster. We found that the line emission is not sensitive to this effect because it is brighter than the continuum, so the fluctuations are not noticeable.

Finally, the output images and cubes created with LIME were passed through the ALMA instrumental response using version 4.4.0 of CASA \citep{McMullin07}. The task `simobserve' was first used to generate visibilities from the model images. The array configuration, integration time, date, hour angle, and precipitable water vapor were all set to properly emulate the observing conditions of the data. The task `clean' was then used to create the ALMA-simulated images from the model visibilities. The simulated and observed images have noise levels and beam sizes matching each other within a few percent. 

\subsection{Iterative building of global model} \label{sec:model_building}

In this section, we summarize the order in which the global model was tailored and the motivation for its specific features. Section \ref{sec:Model_parameters} goes deeper into the determination of the free parameters of the local models.

\begin{figure*}  
\centering
\subfigure[Spectra from local grid, $K=4$]{\includegraphics[scale=0.45]{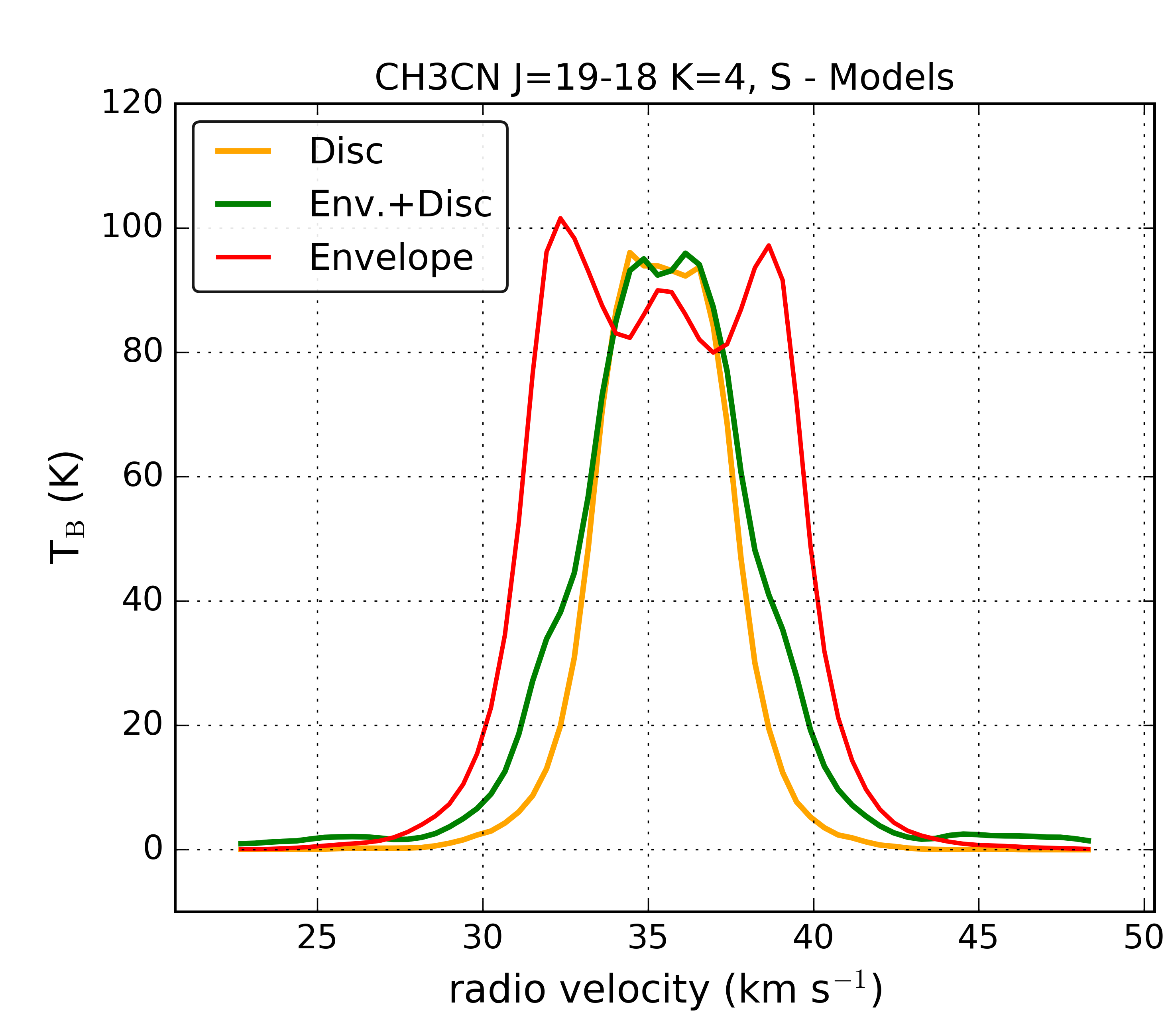}}
\subfigure[Spectra from local grid, $K=8$]
{\includegraphics[scale=0.45]{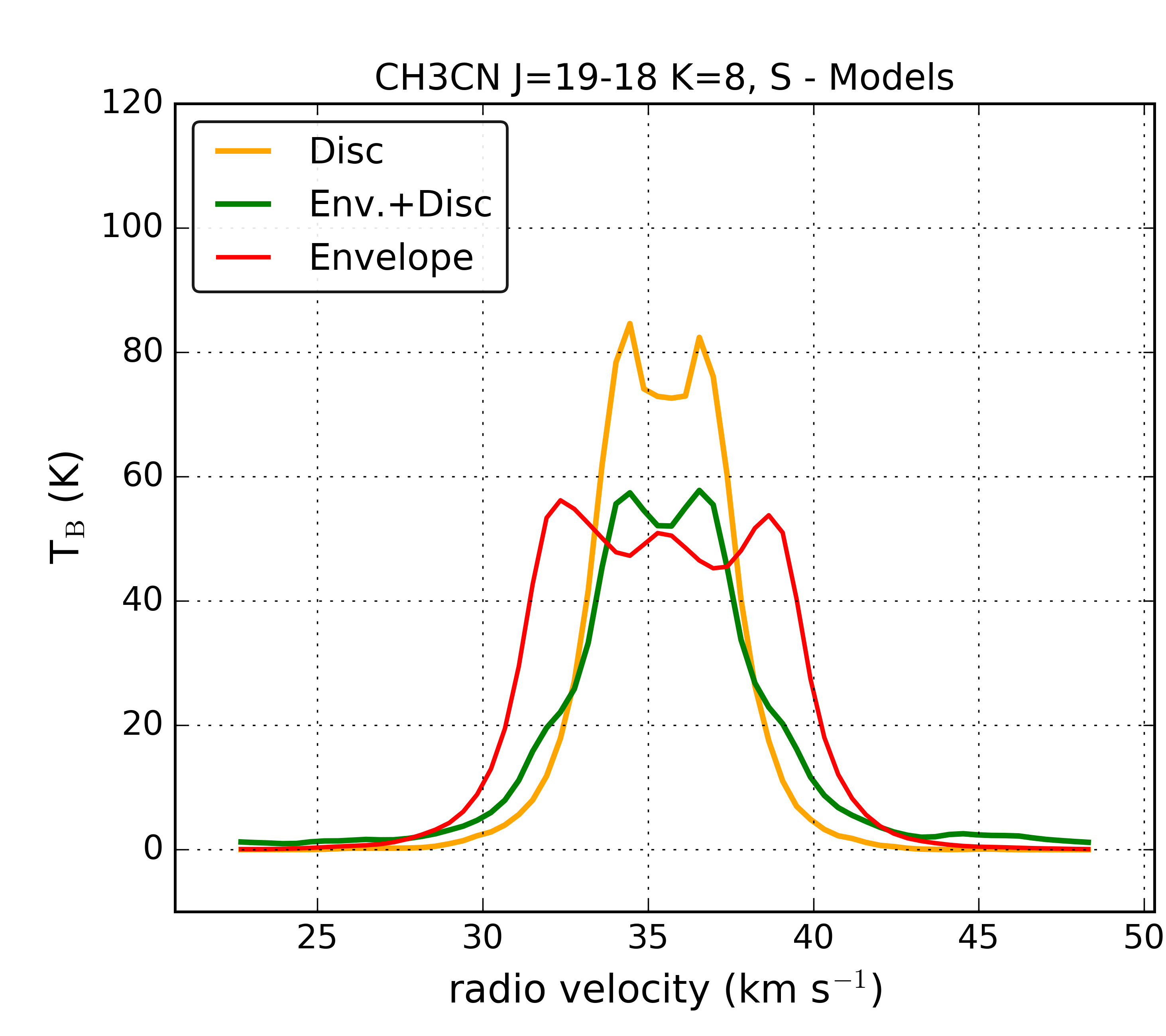}}
\subfigure[Spectra from global grid, $K=4$]
{\includegraphics[scale=0.45]{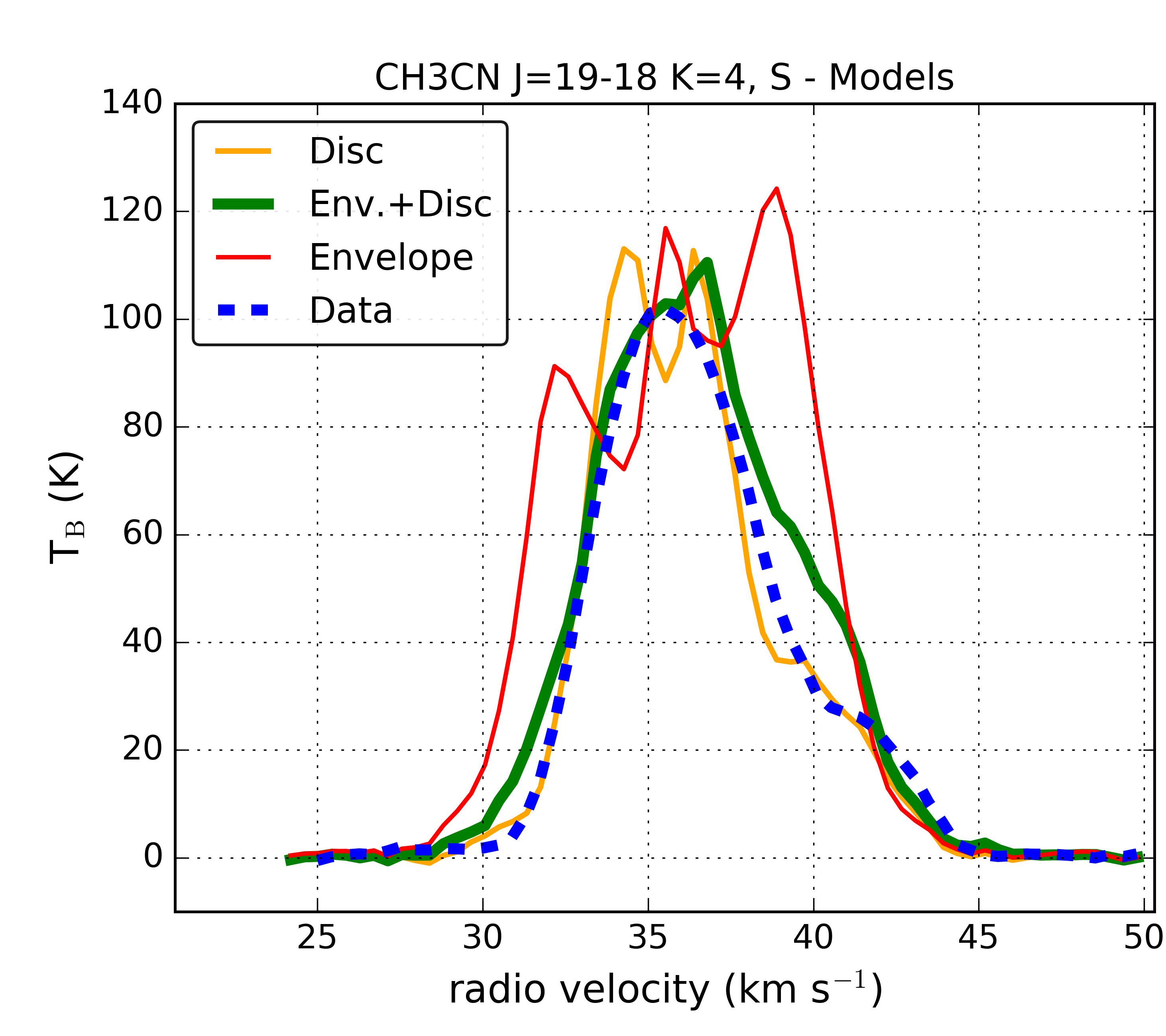}}
\subfigure[Spectra from global grid, $K=8$]
{\includegraphics[scale=0.45]{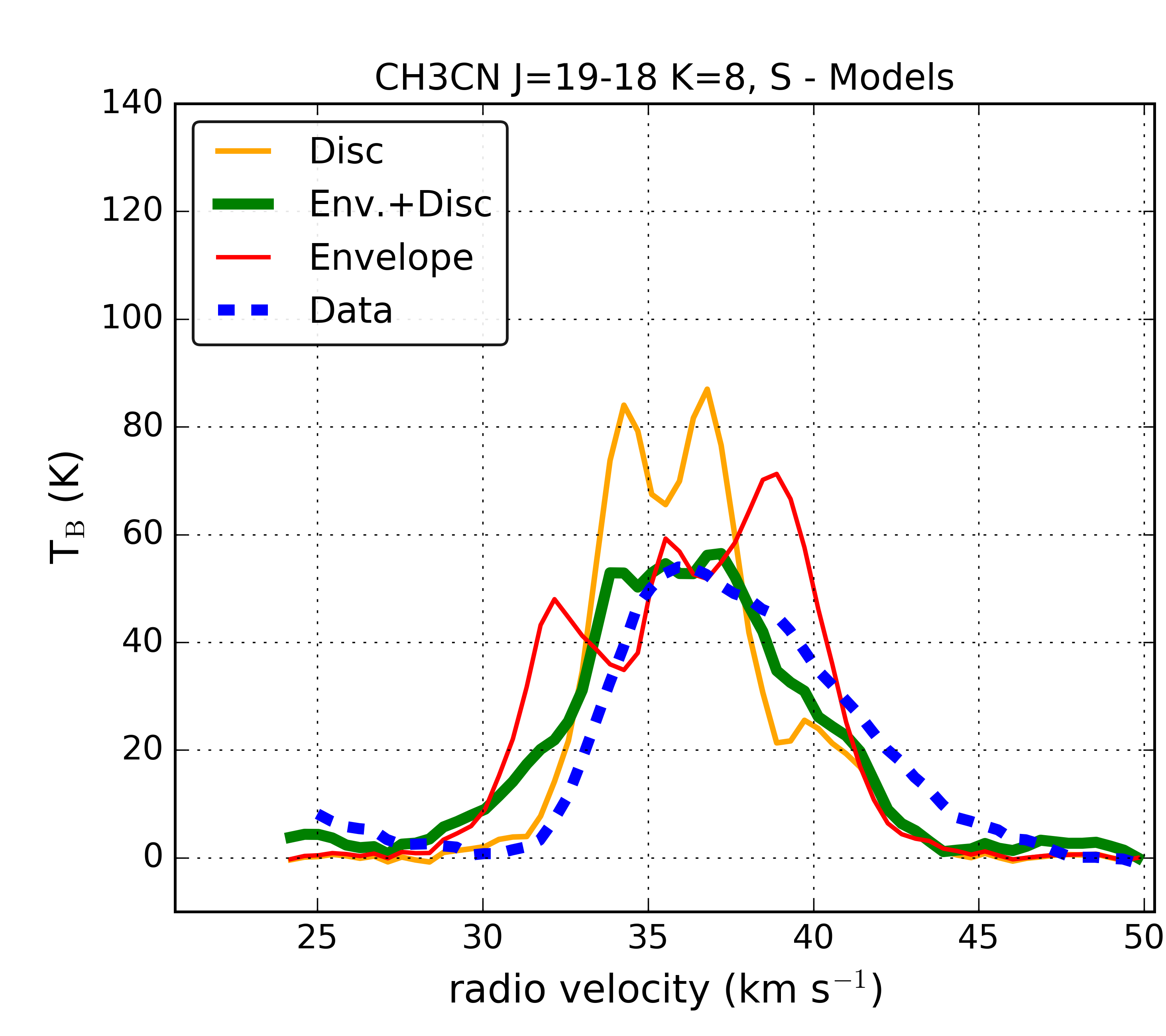}}
\caption{CH$_3$CN $J=19-18$, $K=4$ (left) and $K=8$ (right) spectra for different isolated (top row) and global (bottom row) models for source S. The scenarios are: disc-only (yellow line), envelope-only (red line), and disc+envelope (green line). The dashed lines in the bottom panels show the ALMA data.
The rms noise levels in the spectra are $\sim 1$ K (see Section \ref{sec:observations}.)
}
\label{fig:S_models}.
\end{figure*}

We started modelling compact source Main with a single massive envelope, but in the end a disc embedded within an envelope was a better match to the data. 
Then, we proceeded to model compact sources S, SE, E, and Ridge. Each source was modelled separately in the same way as Main, in its own local grid. The top row of Figure \ref{fig:S_models} illustrates the effects of changing the model prescription for compact source S on its CH$_3$CN line profiles, modelled in isolation. More details can be found in Section \ref{sec:compact_K4}.

The next step was to construct the global grid (Section \ref{sec:GlobalGrid}) and allocate the compact sources within it. 
The first global grid contained only compact sources. 
This global model was compared to the observations, then 
the small inter-source filaments were defined and assigned to a second global grid. 
Small adjustments to the properties of the compact sources and small filaments were done by physically motivated trial and error to better match this second global grid to the observational data. 
The influence of the global environment on the line profiles of source S can be seen in the bottom row of Figure \ref{fig:S_models}.

The final addition to the global grid was the spiral-like filament feature. The assumption of a parabolic orbit with focus at the center of mass of MM1 was readily a good approximation. Then, the 3D orientation and density of the parabola were adjusted to match the observed velocity gradient and fluxes. This filament was initially considered to be static, but a radial collapse component was added to better match its observed velocity dispersion and morphology in the channel maps (see Figure \ref{fig:channels}).

Finally, the newly proposed compact sources MNE and SNW were added to reproduce finer features in the images. Their inclusion helped to reproduce the extended heating and velocity dispersion toward the northeast of Main and the northwest of South, as well as the secondary features around Main and S (more details below). In the resulting global grid, additional small adjustments to each component were tried until we were satisfied with the match between the global model and observations. We reiterate that the models are not best fits to the data, but a physically motivated representation that matches the  observations reasonably well \citep[e.g.,][]{Schmiedeke16}.

\subsection{Determination of model parameters} \label{sec:Model_parameters}

Tables \ref{table:1} and \ref{table:2} list the free and derived model parameters for the compact sources (discs and envelopes) and filaments, respectively. In this section, we describe how the values of those free parameters were determined. We emphasize that our method was intuition-guided trial and error. To find a good set of parameters we iterated over different model versions mainly comparing their CH$_3$CN $K=4$ and 0.8 mm continuum to the data, whereas the $K=8$ line and 1.3 mm continuum were just used as a-posteriori checks.

Although the central stellar mass $M_\star$ is not the unique parameter that affects the line-width, it is the most important (see equations \ref{eq:3} - \ref{eq:5} and \ref{eq:7}). For a first estimation of the stellar mass of the compact sources, we first inspected the channel maps around their central positions as defined from the continuum peaks. In some compact sources, a velocity gradient was clearly discernible (S, SE, Ridge), while in some others it was not (Main, MNE, SNW), probably due to the confusion with neighboring emission. For the sources with clear velocity gradients, we extracted their spectra averaged over the relevant apertures and made Gaussian fits to obtain estimations of the projected rotation velocity of the gas ($\Delta v = 0.5$FWHM), while estimating the disc radius $r$ from measuring the angular separation between the blue- and red-shifted emission lobes. Using these parameters, we made a first estimation of the central mass assuming Keplerian rotation as in $M_\star = \Delta v^2 r / G$. For the sources without clear velocity gradients in the channel maps, we estimated their stellar mass by matching the FWHM of the modelled CH$_3$CN line with the data in the central beam of each. For the case of Main, we started with the 13 $M_\odot$ estimation from \cite{Maud+17}, and reduced it down to 7 $M_\odot$ due to our interpretation of multiplicity (see Section \ref{sec:compact_K4} for more details).

After their first estimation, the central masses were slightly varied to better match the data. The inclination $i$ with respect to the line of sight is also important and is the second preferred parameter that we vary to adjust line-widths. For Main we used the previous estimate based on models by \cite{deWit10}, whereas for the other sources we started with the assumption of $i=45^\circ$ and varied it until we achieved a line-width that matched the observed. 

\begin{figure*}  
\centering
\subfigure[$\theta_h=0^{\rm o}$ \label{fig:6a}]
{\includegraphics[scale=0.5]{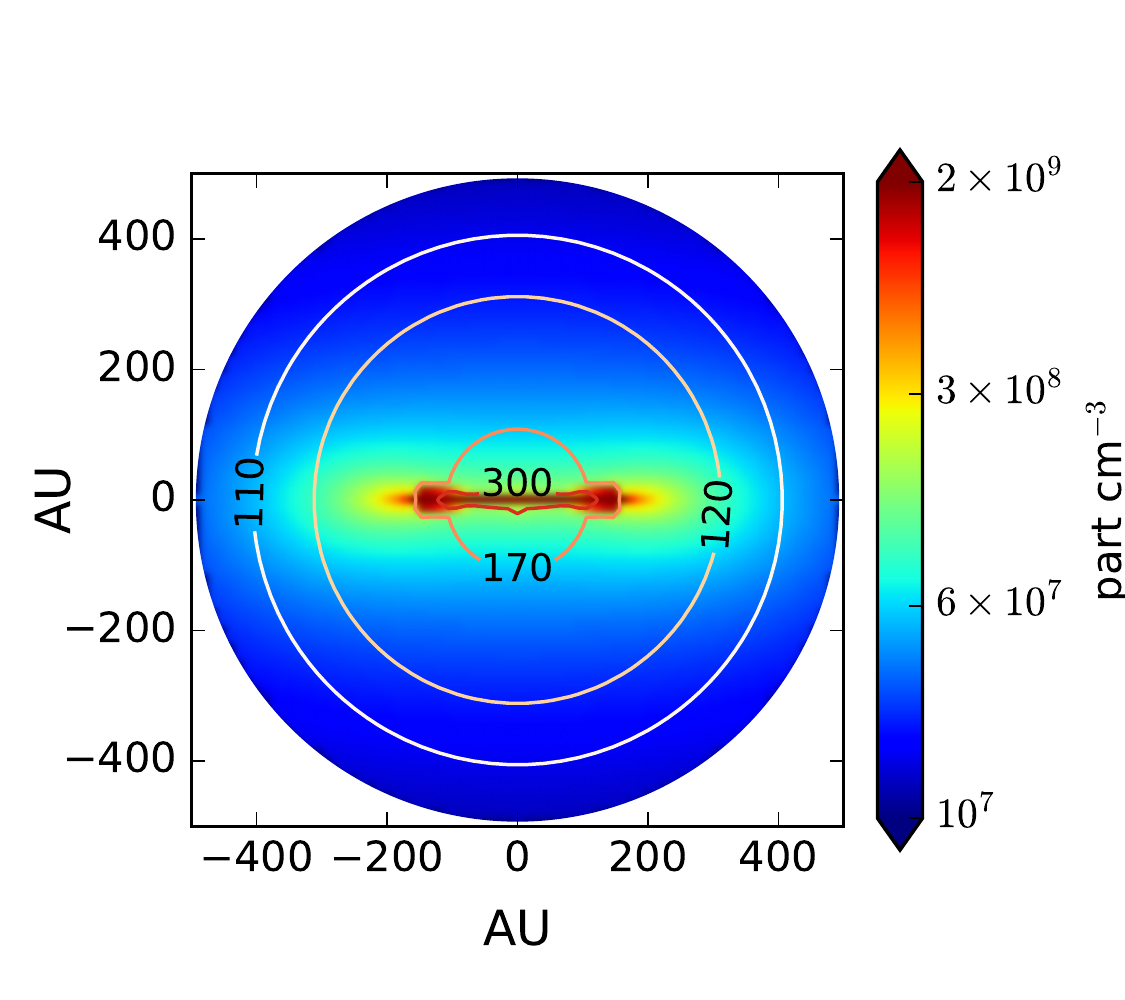}} 
\subfigure[$\theta_h=20^{\rm o}$ \label{fig:6b}]{\includegraphics[scale=0.5]{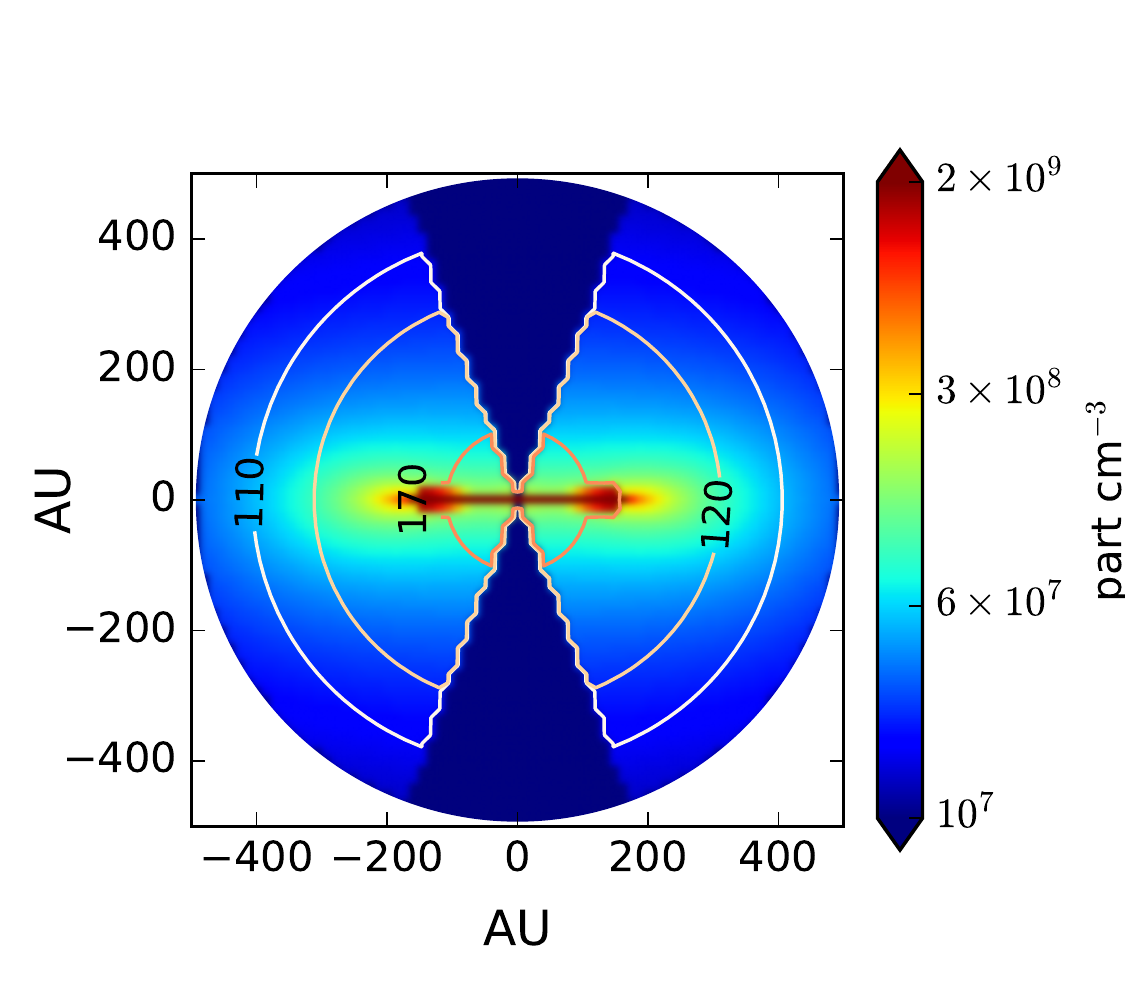}}
\subfigure[$\theta_h=40^{\rm o}$ \label{fig:6c}]
{\includegraphics[scale=0.5]{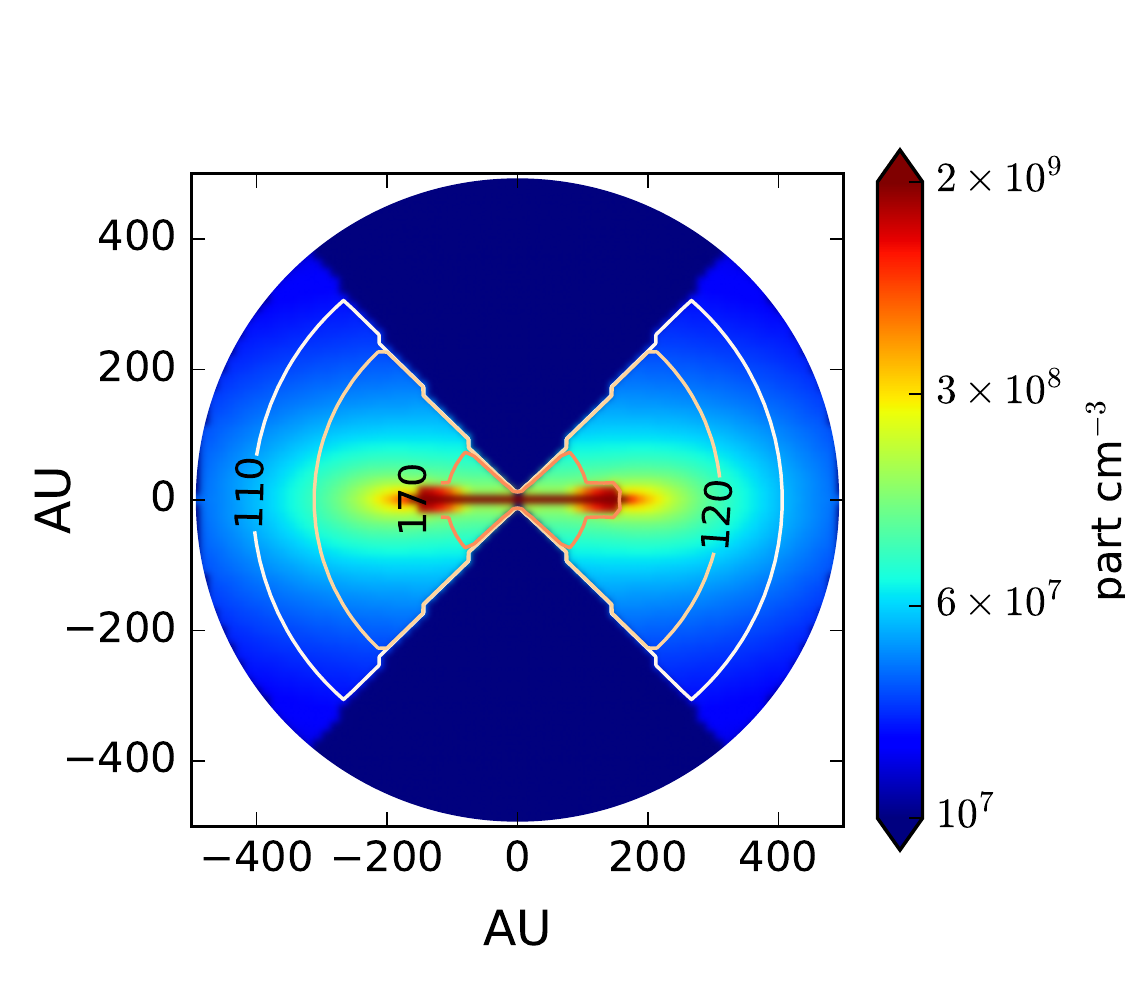}} 
\subfigure[$\theta_h=60^{\rm o}$ \label{fig:6d}]{\includegraphics[scale=0.5]{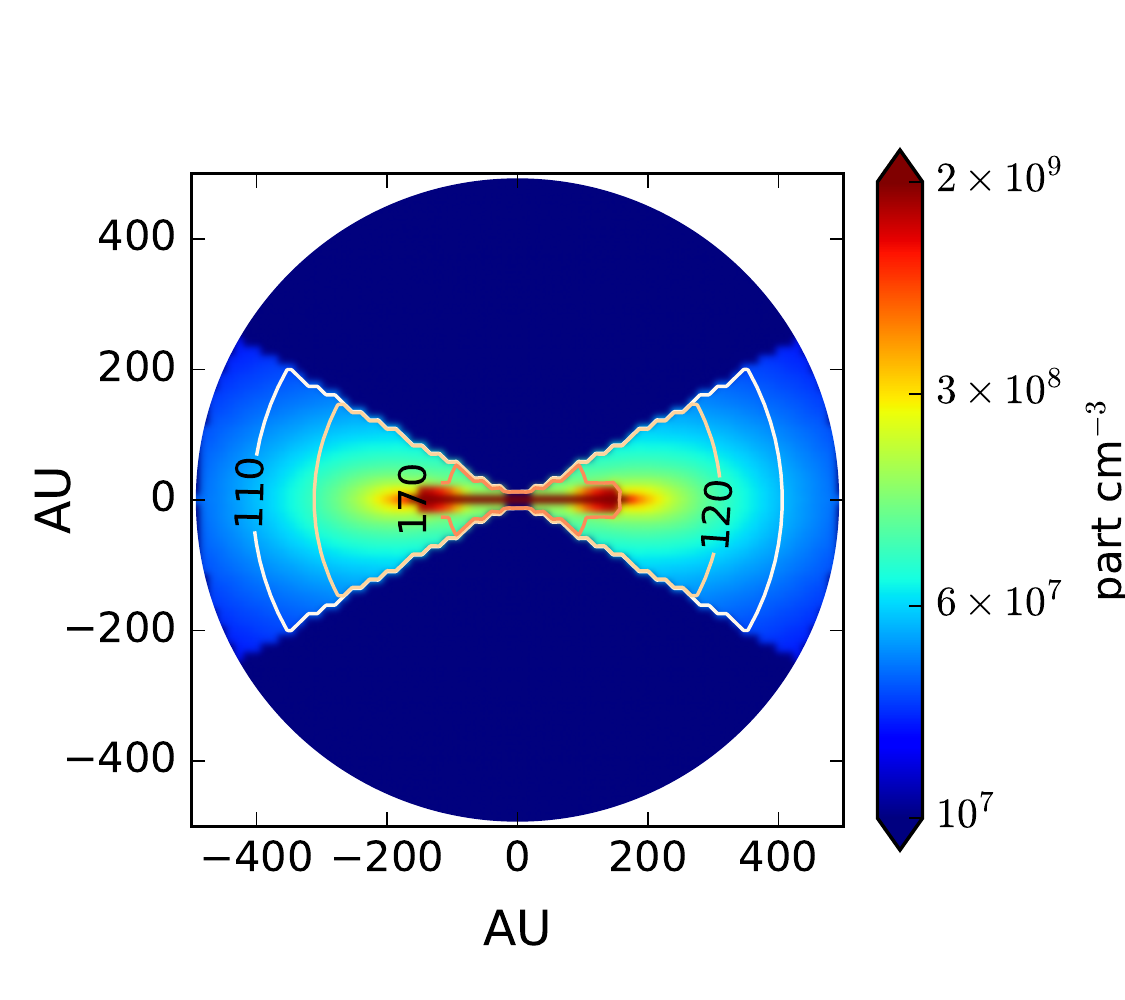}}
\subfigure[Spectra \label{fig:6e}]{\includegraphics[scale=0.47]{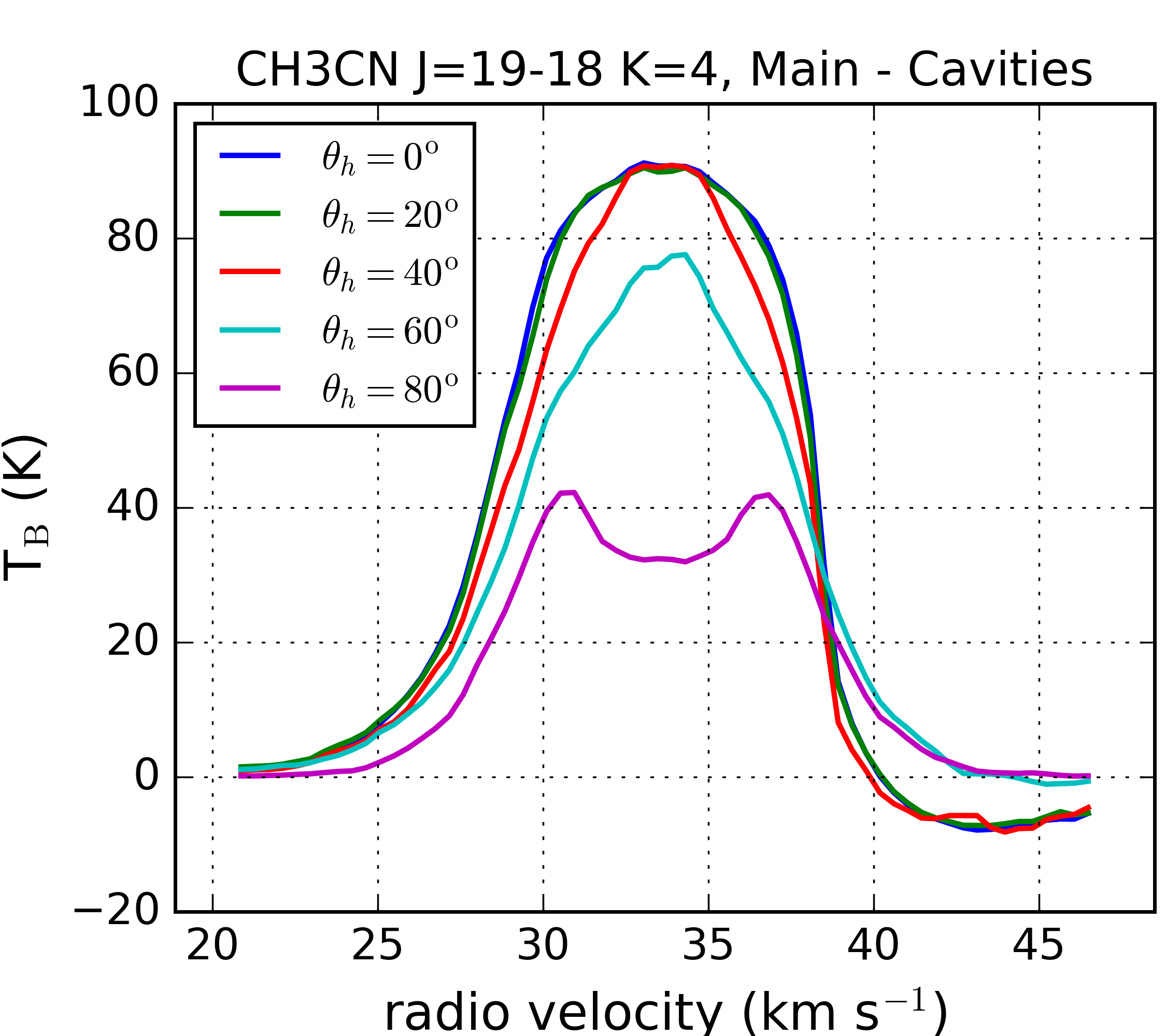}}
\subfigure[$\theta_h=80^{\rm o}$ \label{fig:6f}]
{\includegraphics[scale=0.5]{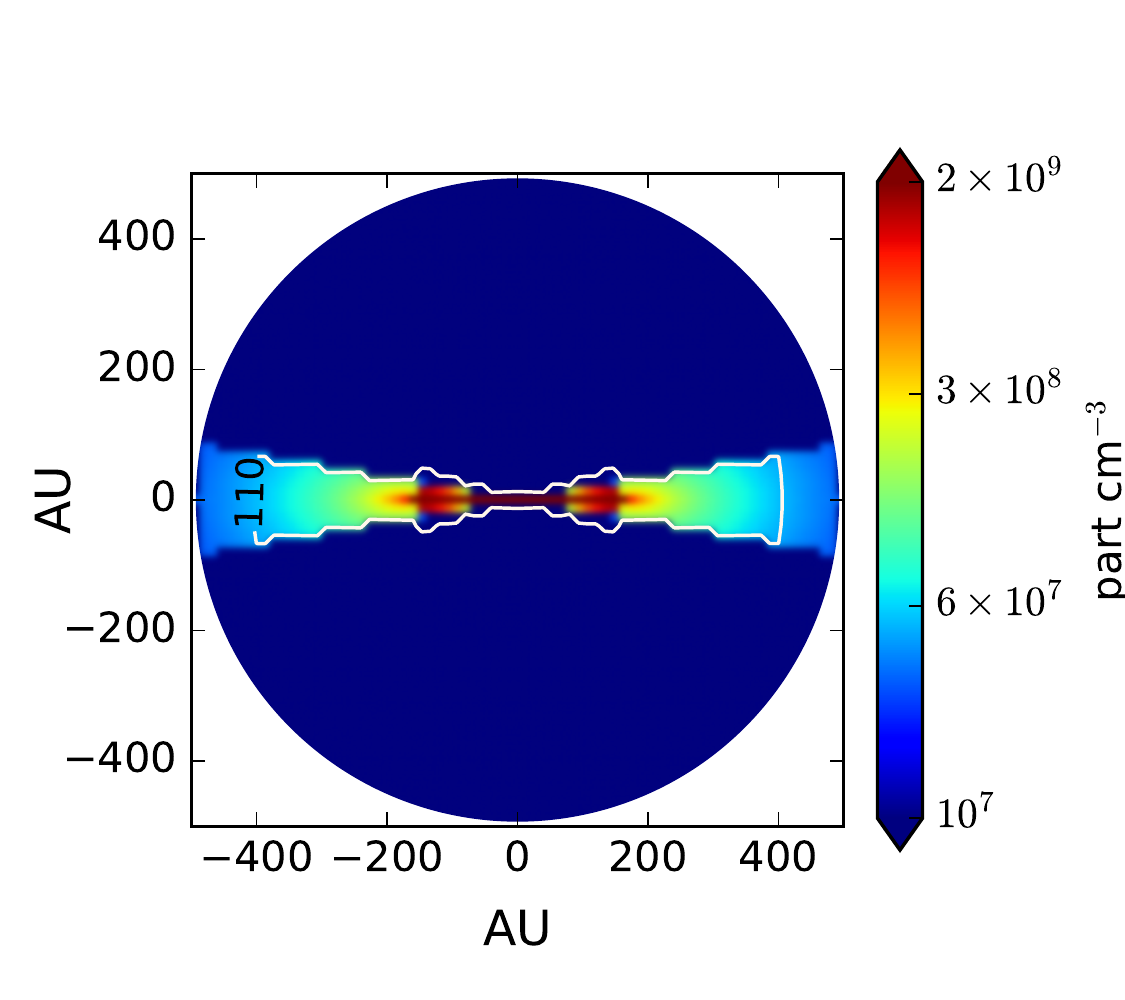}} 
\caption{Effect of varying the cavity opening angle on the spectra of compact source Main modelled in isolation. The upper row shows from left to right: $\theta_h=0^\circ,20^\circ,40^\circ$. The bottom row shows $\theta_h=60^\circ$ (left) and $\theta_h=80^\circ$ (right). The middle panel of the bottom row shows the resulting CH$_3$CN $J=19-18$, $K=4$ spectra. Density (colors) and temperature (contours) profiles were also included in the subplots. The cross sections show edge-on models ($i=90^\circ$). The spectra come from models with inclination $i=45^\circ$, the same as Main in the model. It is clear that the two-peaked profile is only seen with cavities wider than $60^\circ$ when the disc dominates the emission.} 
\label{fig:cavity}
\end{figure*}

A cavity with opening half-angle $\theta_h$ was included in the models of Main and S. Observations show that MM1 has at least one molecular outflow centered in Main \citep{GM10,Davies10}. It is not clear whether S also drives an outflow, but we incorporated a cavity in its model given that it is the second most massive source. The cavity was used mainly to refine the line profile of these sources but also to construct a more realistic model of their inner regions. We varied $\theta_h$ from $0^\circ$ to $80^\circ$ in steps of $20^\circ$ (see Figure \ref{fig:cavity}). 

The mass accretion rate $\dot{M}$ was varied as a free parameter in order to scale up or down the density of the compact sources, with a corresponding effect on the line and continuum intensity levels. From equation \ref{eq:9}, the normalization constant $\rho_{\rm e_0}$ of the envelope density is directly proportional to $\dot{M}$. Note that as a consequence, this is also true for the disc density (see equation $\ref{eq:6b}$). Since Main is a high-mass protostellar object, we used $\dot{M}$ in the range $10^{-4}$ to $10^{-3}$ $M_\odot$ yr$^{-1}$ \citep[e.g.,][]{Osorio09,ZY07}. For the lower-mass sources we tested values in the range $10^{-6}$ to $10^{-5}$ $M_\odot$ yr$^{-1}$.

The normalization of the envelope temperature and the disc temperature factor $B_T$ were set such that the resultant density weighted mean temperature of the compact sources was consistent with the temperature map presented in \cite{Maud+17}. Similarly, the disc density ratio $A_\rho$ is used to calibrate the density-weighted mean temperature, as well as the continuum and line intensities, with the observational data. We restrict both parameters $B_T$ and $A_\rho$ to values $\sim 5$ to $20$, in order to not exaggerate the relative importance of the disc with respect to the envelope. 

The centrifugal radius of the envelope $R_\mathrm{d}$ was chosen to be the same disc radius previously defined in this section. The line emission of MNE and SNW is marginally (un)resolved, thus, we restricted $R_\mathrm{d}$ to be larger than half of the envelope size. The emission in Main is quite compact, so we set $R_\mathrm{d}$ from the line-width and continuum intensity in its central beam. The position angle $PA$ in the plane of the sky was also defined from the same restrictions as $R_\mathrm{d}$. For Main, previous observations and modelling provided a good first estimate \citep{deWit10,Davies10}.

For the CH$_3$CN abundance with respect to H$_2$, N$_{\rm CH_3CN/H_2}$, we tested values from 10$^{-9}$ to 10$^{-7}$, in agreement with determinations using interferometric observations in massive star formation regions \citep[e.g.,][]{Wilner94,Remijan04,GM09}. 
We started with N$_{\rm CH_3CN/H_2} $=10$^{-9}$ and increased it in order to adjust the line emission once the correct continuum flux was achieved.

The systemic velocity $v_{\rm sys}$ of the compact sources was set to be the velocity of the line peak in the data. For cases like Main where the $K=4$ CH$_3$CN line was optically thick at the center, the $K=8$ line was also considered to refine $v_{\rm sys}$.

The dominating mass is the mass responsible for the gravitational field that determines the gas velocity in the filaments (see equations \ref{eq:11} and \ref{eq:12}). For the large parabolic filament, the dominating mass is set to the total model mass contained in the central region of MM1, \ie, stellar $+$ gas mass of Main $+$ S $+$ MNE $+$ SNW $+$ the gaseous mass in their vicinity. For the cylindrical filaments converging to Main or MNE, the dominating mass is the mass of Main. For the rest of the small filaments, it is the mass of the most massive of the two sources at the extremes of the cylinder.

The temperature in each of the filaments was set to be homogeneous, and consistent with the determination in \cite{Maud+17}. The density for each filament, homogeneous as well, was used mainly to control the continuum intensity, whereas the CH$_3$CN abundance was used to further adjust the line emission. We tested H$_2$ particle densities from 10$^6$ to 10$^8$ cm$^{-3}$ and abundances from 10$^{-9}$ to 10$^{-7}$.

The length of the filaments was estimated from the spatial configuration of the compact sources in the data. We assume that the cylinders have the same depth in the line of sight as projected length in the plane of the sky (see Section \ref{sec:acc_filaments-results} for further details). The cylindrical radii were estimated directly from the apparent angular width of the filaments in the continuum and channel maps.

\setlength{\tabcolsep}{7pt} 

\begin{table*} 
\centering
{\renewcommand{\arraystretch}{1.15}
\begin{tabular}{@{} l *7c  @{} }
\toprule
 \multicolumn{1}{c}{Parameter}    & Main  & MNE  & S  & SNW  & SE  & E  & Ridge \\ 
\midrule
 Stellar mass ($M_*$) [M$_{\odot}$] & 7.0 & 0.6 & 2.8 & 0.6 & 0.9 & -- & 0.9\\ 
  Mass accretion rate ($\dot{M}$) [M$_{\odot}$ yr$^{-1}$] & 4.0 $\times$ 10$^{-4}$ & 1.3 $\times$ 10$^{-5}$ & 1.0 $\times$ 10$^{-5}$ & 2.8 $\times$ 10$^{-5}$ & 1.1 $\times$ 10$^{-5}$ & -- & 1.0 $\times$ 10$^{-5}$\\  
 Env. temp. at 10 AU ($T_{10_{\rm env}}$) [K] & 375 & 2500 & 1875 & 1875 & 500 & -- & 250\\ 
 Centrifugal radius ($\Rd$) [AU] & 152 & 302 & 362 & 202 & 264 & -- & 164 \\ 
 Cavity half opening angle ($\theta_h$) [deg.] & 40 & -- & 20 & -- & -- & -- & -- \\
 Disc density ratio ($A_\rho$) & 24.1 & -- & 7.5 & -- & 11.9 & -- & 12.6\\  
 Disc temperature factor ($B_T$) & 5.0 & -- & 15.0 & -- & 11.3 & -- & 3.8\\
 Abundance (N${_{\rm CH3CN}}$/N${_{\rm {H_2}}}$) & 1.8 $\times$ 10$^{-7}$ & 1.3 $\times$ 10$^{-7}$ & 6.0 $\times$ 10$^{-7}$ & 3.8 $\times$ 10$^{-8}$ & 1.0 $\times$ 10$^{-7}$ & 1.8 $\times$ 10$^{-8}$ & 6.7 $\times$ 10$^{-8}$\\
 Position Angle (\textit{PA}) [deg.] & -23 & 23 & 15 & 45 & 10 & -- & 5 \\ 
 Inclination (\textit{i}) [deg.] & 45 & 40 & 30 & 40 & 45 & -- & 45\\ 
 Systemic velocity ($v_{\rm {sys}}$) [km s$^{-1}$] & 33.7 & 41.7 & 35.5 & 41.2 & 38.1 & 38.7 & 38.2\\ 
 Local grid radius [AU] & 500 & 400 & 400 & 300 & 500 & 400 & 300 \\ 
 \midrule
 Stellar radius ($R_*$) [R$_{\odot}$] & 30.2 & 3.8 & 2.3 & 5.2 & 0.9 & -- & 0.9\\  
 Mean temperature$^a$ [K] & 990 & 832 & 806 & 695 & 413 & 140 & 201\\
 Envelope density at $\Rd$ ($\rho_{\rm e_0}$) [cm$^{-3}$] & 1.8 $\times$ 10$^{8}$ & 7.2 $\times$ 10$^{6}$ & 2.0 $\times$ 10$^{6}$ & 2.8 $\times$ 10$^{7}$ & 6.2 $\times$ 10$^{6}$ & 1.3 $\times$ 10$^{7}$ $^b$ & 1.2 $\times$ 10$^{7}$\\ 
 Envelope mass ($M_{\rm env}$) [M$_{\odot}$] & 0.122 & 0.009 & 0.003 & 0.014 & 0.010 & 0.039 & 0.005\\ 
 Disc mass ($M_{\rm disc}$) [M$_{\odot}$] & 0.271 & -- & 0.011 & -- & 0.028 & -- & 0.012 \\ 
 Total mass of gas ($M_{\rm gas}$) [M$_{\odot}$] & 0.393 & 0.009 & 0.014 & 0.014 & 0.038 & 0.039 & 0.017 \\
 

 \bottomrule
 \end{tabular}
 
  \caption{{\it Top:} Free parameters for compact sources. {\it Bottom:} Derived properties from model output. $^a$: Density weighted mean temperature. $^b$: Mean density in E.}
  \label{table:1}
 }
\end{table*}

\setlength{\tabcolsep}{7pt} 

\begin{table*} 
\centering
{\renewcommand{\arraystretch}{1.15}
\begin{tabular}{@{} l *6c  @{} }
\toprule
 \multicolumn{1}{c}{Parameter}    & Spiral  & E $\rightarrow$ MNE & E $\rightarrow$ SE & SE $\rightarrow$ MNE & SE $\rightarrow$ Main & SE $\rightarrow$ S \\ 
\midrule
 Dominating mass ($M_{\rm c}$; $M_{\star^>}$) [M$_{\odot}$] & 12.0 & 7.0 & 0.9 & 7.0 & 7.0 & 2.8  \\
 Filament density ($\rho_0$) [cm$^{-3}$] & 8.0 $\times$ 10$^{7}$ & 4.3 $\times$ 10$^{8}$ & 8.0 $\times$ 10$^{7}$ & 8.3 $\times$ 10$^{6}$ & 8.3 $\times$ 10$^{6}$ & 3.8 $\times$ 10$^{7}$ \\ 
Filament temperature [K] & 200 & 250 & 250 & 250 & 250 & 250 \\
Filam. abundance (N${_{\rm CH3CN}}$/N${_{\rm {H_2}}}$) & 1.5 $\times$ 10$^{-8}$ & 0.8 $\times$ 10$^{-8}$ & 1.0 $\times$ 10$^{-8}$ & 5.0 $\times$ 10$^{-8}$ & 5.0 $\times$ 10$^{-8}$ & 0.7 $\times$ 10$^{-8}$ \\
 Cylindrical radius [AU] & 200 & 150 & 100 & 150 & 150 & 150 \\
 Length [AU] & 7240 & 1300 & 870 & 2060 & 1540 & 2430 \\ 
 \midrule
 Mean tangential velocity [km s$^{-1}$] & 4.0 & 3.9 & 1.1 & 3.1 & 3.1 & 1.5 \\ 
 Mass inflow rate [10$^{-5}$ M$_{\odot}$ yr$^{-1}$] & 4.75 & 2.79 & 0.32 & 0.22 & 0.22 & 0.48 \\
 Mass of gas ($M_{\rm gas}$) [M$_{\odot}$] & 0.410 & 0.045 & 0.012 & 0.007 & 0.005 & 0.036 \\
 \bottomrule
 \end{tabular}

 \caption{Same as Table \ref{table:1} but for the filamentary structures.} \label{table:2}
 }
\end{table*}

\section{Results} \label{sec:results}

\subsection{CH3CN J=19--18, K=4} \label{sec:K4}

\subsubsection{Compact sources} \label{sec:compact_K4}

\begin{figure*}  
 \centering
 \includegraphics[width=1.0\textwidth]{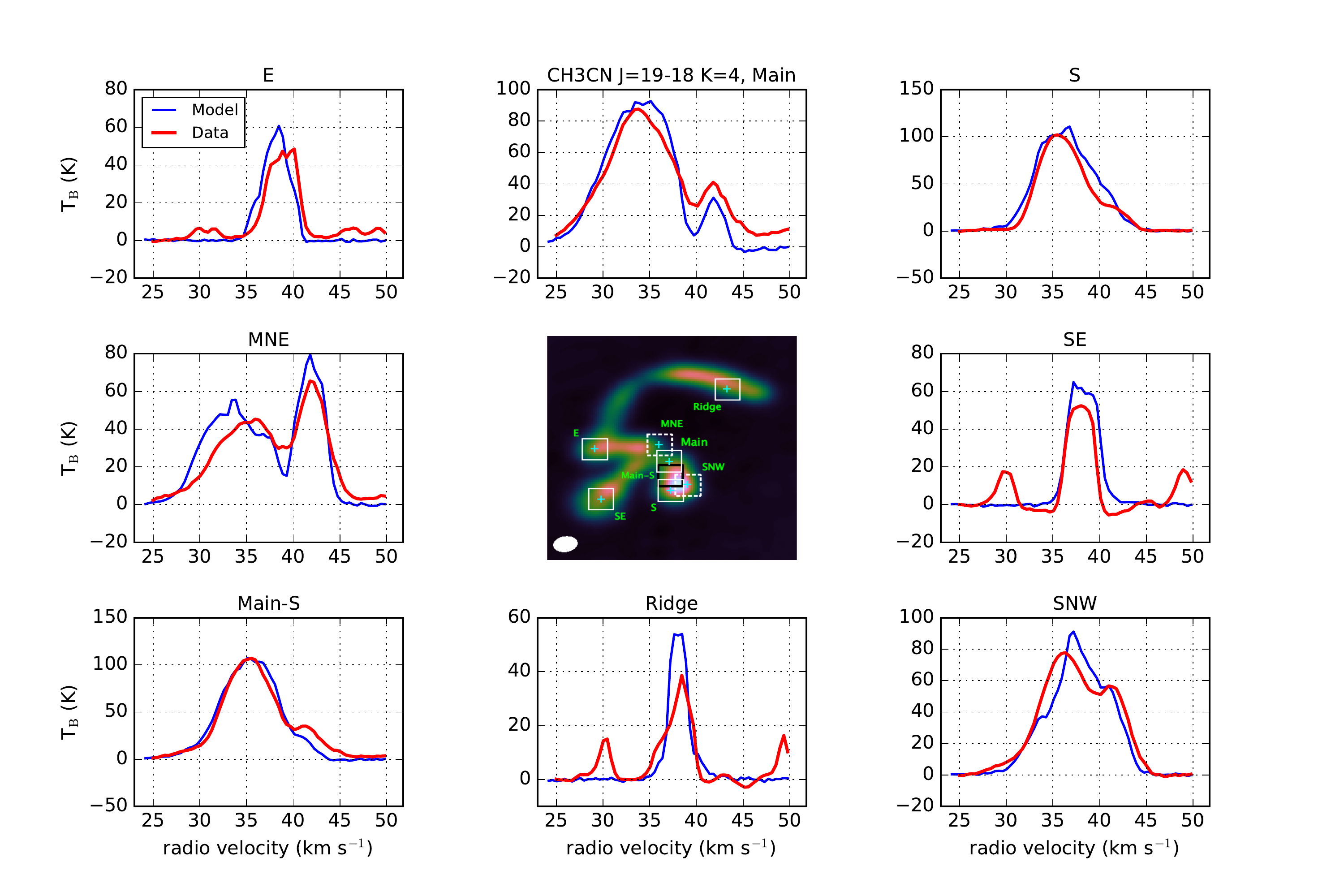} 
 \vspace{-0.5cm}
 \caption{CH$_3$CN $J=19-18$ $K=4$ spectra of the modelled compact sources compared with the data. The header on each panel indicates the region of analysis. The central panel is a snapshot of the velocity channel $v = 38.05$ km s$^{-1}$, where the $0.2\times0.2$ arcsec integration apertures around compact sources are marked in white. The dashed line apertures are centered in the new compact sources proposed by the model (MNE and SNW). The black square shows an extra aperture of interest in between compact sources Main and S, labeled as Main-S in the spectra. The cyan markers indicate the center of the compact sources included in the model.}  
\label{fig:spectra_sources}
\end{figure*}

To compare the model spectra of the compact sources with those from the ALMA images, in Fig. \ref{fig:spectra_sources} we show the average spectra in squared apertures of 0.2 arcsec size (approximately the beam size). The observed lines are generally asymmetric, and in most cases there are secondary velocity features besides the principal line peaks. The presence of such features in apertures already as small as 480 au, as well as the absence of pure two-peaked line profiles, warns against interpreting the compact sources only as Keplerian, rotationally supported discs. Nearby companions, envelopes, and filamentary flows can all contribute to the observed spectra. 

The spectrum around compact source {\it Main} consists of a single-peaked line centered at $\sim 34$ km s$^{-1}$, and a secondary, fainter component peaked at $\sim 42$ km s$^{-1}$ (see Fig. \ref{fig:spectra_sources}). In our model, the brightest line peak is dominated by Main, whereas the redshifted, fainter spectral component is contributed by compact source MNE. Our interpretation also reproduces the spectrum around the central position of MNE.

To model Main we first considered the simplest case of a pure Keplerian disc, but such model always produces double-peaked profiles unless an unrealistically high optical depth -- mass -- is used, which consequently also produces line and continuum fluxes that are too high. We concluded that an envelope surrounding the disc is the most natural way to produce the single-peaked profile of the brightest spectral component (see Fig. \ref{fig:spectra_sources}) while matching both the $K=4$ and continuum fluxes. The inclination angle was chosen based on the restrictions available from the IR interferometric and photometric modelling of \cite{deWit10}, who found $i \sim50^\circ$ (we set $i = 45^\circ$ for simplicity). 
Several observations show that Main drives one, or possibly two massive molecular outflows \citep{GM10,Davies10,Maud+15} and possess a cavity on scales of a few $\times100$ au \citep{deWit10}. We included such cavity in the model. Figure \ref{fig:cavity} shows the effect of changing the opening angle of the cavity on the model spectra of Main. An opening angle $\theta=40^\circ$ is a good compromise between the need to wash out the two-peaked line profile (wider cavities do not since for them the model approximates to a pure disc without envelope) and having a line that is not too broad and too bright, as in the case of much narrower cavities. 

Although Main is still the most massive object in the field (the full list of free parameters and derived quantities is in Table \ref{table:1}), we determined a lower central mass and smaller disc size than previous estimates. In previous (sub)mm observations where the entire MM1 core\footnote{We refer to MM1 as a `core' following the convention of using this word for structures of $\sim 0.03$ to 0.2 pc size \citep[e.g.,][]{BerginTafalla07}. The individual discs and envelopes in our model are smaller scale structures within the MM1 core.} was marginally resolved, the kinematics was interpreted as originating from a dynamical mass of 10 to $15~M_\odot$ within a $\sim 1200$ au radius \citep{GM10}. \cite{Maud+17} determined a dynamical mass $\sim 13~M_\odot$ within a 1000 au radius from the current ALMA data. Given that we interpret emission peaks MNE, SNW, and S as {\it separate} YSOs, and also consider several filamentary flows feeding Main, MNE, and S from the east and southeast, our model of Main requires a lower stellar mass ($7~M_\odot$), dynamical mass (stellar+disc+envelope, $7.4~M_\odot$), and disk radius ($\sim 150$ au) than previous estimates. The total stellar+gas mass of our model within a 1000 au radius of the stellar source in Main is $\sim 12~M_\odot$, consistent with previous estimates. The 150 au disc radius that we propose for Main is unresolved by our observations with 400 au resolution, but ALMA long-baseline data will be able to test our hypothesis or discard the existence of a {\it true disc} on scales $<100$ au. 

For compact source \textit{S}, as for the case of Main, the observed profile is not a simple two-peaked line. In this case there is a dominant peak at $\sim 36$ km s$^{-1}$, with a fainter, redshifted shoulder from $\sim 40$ to 44 km s$^{-1}$, which in our model comes from the neighboring compact source SNW (see below).
Figure \ref{fig:S_models} shows a comparison of the CH$_3$CN $J=19-18$, $K=4$ and $K=8$ lines for source S resulting from three model scenarios: an envelope without disc, a disc without envelope, and a disc+envelope. The line profile of the pure envelope, and more prominently, of the pure disc, have the two peaks characteristic of rotation, whereas in the disc+envelope model the peaks are much less noticeable. To reach the desired peak intensity at the $K=4$ transition, we scaled up the density through increasing the mass accretion rate in the different model scenarios. A pure envelope needs to become too optically thick over a wide velocity range, in which case the line is too `square-shaped'. Similarly, a pure disc does not preserve the desired peak intensity at the $K=8$ transition. Regarding to the continuum emission, the pure envelope has low mean and peak intensities ($\sim$ 1/5 compared to the data at 349 GHz), whereas the pure disc generates intensities that are too high by $\times$4. Moreover, only the disc+envelope model shows a line profile similar to the data in both transitions while maintaining the correct continuum intensities. 
Thus, we concluded that a disc+envelope is the best model for source S. We included a cavity in this model too, given that S is the second most massive source.
After spanning the possible range of values for the inclination angle with respect to the line of sight $i$ and the cavity half-opening angle $\theta_h$, we found that $i=30^\circ$ (closer to face-on than to edge-on) and $\theta_h=20^{\circ}$ match well the observations (see Fig. \ref{fig:spectra_sources}). The results are degenerate, but more sensitive to variations of the former parameter.

Two small peaks close to Main and S are reproduced by including two new compact sources, which we label Main NE ({\it MNE}) and South NW ({\it SNW}). The emission from these objects is reproduced by considering two low-mass sources with a central mass of $0.6~M_\odot$ surrounded by a pure Ulrich envelope. The model spectrum of SNW has a significant contribution from the nearby, brighter source S. Similarly, the spectrum of MNE is influenced by the contribution from Main (see Fig. \ref{fig:spectra_sources}).

Compact source {\it SE} is modelled as a disc+envelope system, in a similar way to Main and S. There are no previous constraints on its disc inclination, so we decided to set it to 45$^\circ$. The model reproduces the principal line peak at $\sim 38$ km s$^{-1}$, but does not reproduce the fainter blue- and red-shifted components (Fig. \ref{fig:spectra_sources}), which likely arise from contamination of neighboring molecular lines. This contamination is also apparent at similar velocity ranges in the spectra around compact sources E and Ridge, all of them related to the larger spiral-like filament.  

Compact source {\it E} does not have any evidence of velocity gradients in the observational cubes. Therefore, we decided to model it as a sphere with a power-law density profile and a random velocity distribution. The velocity dispersion was chosen to reproduce the observed linewidth. This source probably represents a younger, lower-mass, prestellar core.  

Compact source {\it Ridge} is embedded within the spiral-like filament before it reaches the crowded central part of MM1. Its relatively high brightness and velocity dispersion motivated us to model it as a disc-envelope system rather than just a turbulent sphere as was the case for source E. The model line is brighter and narrower than in the data, but still matches them reasonably. 

We also show in Fig. \ref{fig:spectra_sources} the zone between Main and S, labeled as Main-S. In the model, the spectrum of Main-S has contributions from three compact sources: Main, S and SNW. The good match illustrates that our model is a reasonable approximation of the observed system.  Table \ref{table:1} lists the final model parameters for the compact sources. 

\subsubsection{Accretion filaments} \label{sec:acc_filaments-results}

\begin{figure*}  
 \centering
 \includegraphics[width= 1.0\textwidth]{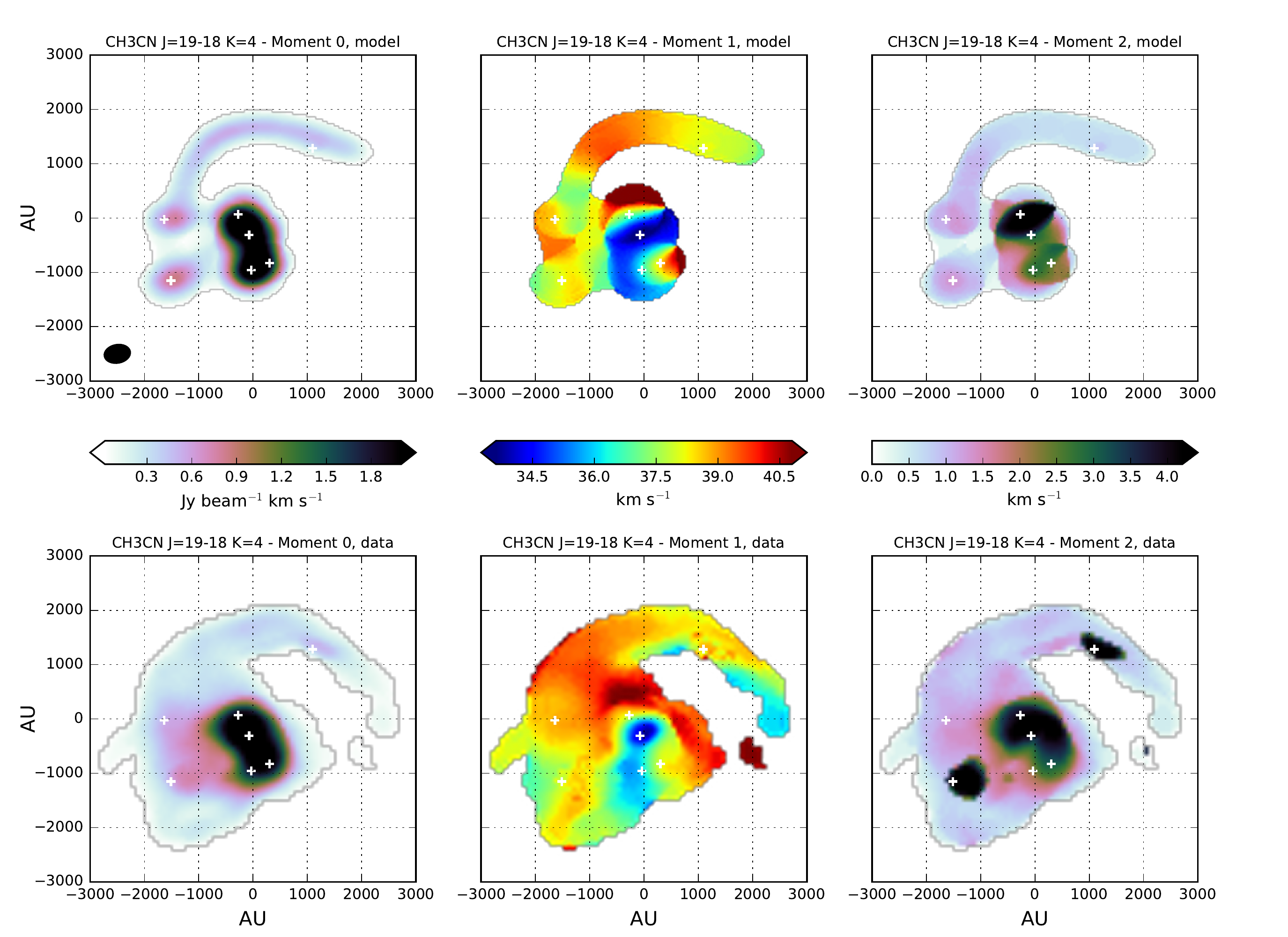} 
 \caption{{\it Top row:} intensity moments for the model CH$_3$CN $J=19-18$, $K=4$ line. {\it Left:} velocity-integrated intensity (moment 0). {\it Center:} intensity-weighted mean velocity (moment 1) integrated between 25.0 and 49.8 km s$^{-1}$. {\it Right:} intensity-weighted velocity dispersion ($\sigma$, moment 2) over the same integration range as the moment 1. {\it Bottom row:} same as top row but for the ALMA data. Cells with intensity below 10\% of the peak were masked out.  
The compact sources included in the model are marked by crosses (+). The beam is shown in the lower left corner of top-left panel.}  
\label{fig:moments}
\end{figure*}

Two types of accretion filaments are included in our global model of W33A MM1 (see Section \ref{sec:filaments}): a larger spiral-like filament feeding MM1 from the outside, and smaller, straight filaments joining pairs of compact sources. Table \ref{table:2} lists the parameters of the selected models.

Figure \ref{fig:moments} shows the moment maps (velocity integrated intensity, intensity-weighted mean velocity, and intensity weighted velocity dispersion) of the model and the observational data. It is seen that the \textit{spiral-like `feeding' filament} model is a good description of the observations. This filament approaches the central part of MM1 coming from the near, north-west side of the observer, and moves toward the east and away from the observer, to finally turn toward the observer while merging with MM1 close to compact source E.  We emphasize that the model is physically motivated, since the velocity field was calculated assuming test particles that approach from the infinite at rest and follow a parabolic trajectory in which the mass of the central region of MM1 resides at its focus. This simple prescription naturally reproduces the blueshifted-redshifted-blueshifted pattern of the line-of-sight velocity across the filament as projected onto the plane of the sky. The real filament seems somewhat more closed and extended than the model in its far end, something that a purely parabolic trajectory cannot reproduce. 
In spite of being quite warm (200 K), an extra velocity component was needed to reproduce the observed velocity dispersion along the trajectory of this filament.
We therefore implemented the reasonable assumption that the filament has radial, transonic collapse. With $\gamma = 7/5$ (diatomic molecules) we set a radial infall velocity $v_{in} = 1.5c_{\rm s} = 1.6$ km s$^{-1}$ (see Section \ref{sec:filaments}), something in between the subsonic and supersonic collapse observed toward low- and high-mass star forming cores, respectively \citep[e.g.,][]{Keto15,GM09}. Adding this radial component increased the velocity dispersion to levels close to the observed, although still slightly below. Figure \ref{fig:spectra_regions} shows a comparison of the model and observed spectra in an aperture containing the entire spiral-like filament. 

The small (length $\sim 10^3$ au) cylindrical filaments are required to reproduce the elongated emission joints observed between compact sources in the line emission maps and the (sub)mm continuum (see Section \ref{continuum}).  Again, the models are  physically plausible since we consider that the gas follows Newtonian dynamics and go from the less massive object to the more massive one. Their length in the line-of-sight direction is considered to be the same as their projected size in the plane of the sky. Two such flows go from compact source E to SE and MNE, and three more go from compact source SE to Main, S, and MNE. Table \ref{table:2} lists the selected parameters of these five filaments. The existence of the two filaments that cross diagonally (SE$\rightarrow$Main and SE$\rightarrow$MNE) is not clear, but including them helped to reproduce line emission extending toward the northwest of compact source SE. Figures \ref{fig:moments} and \ref{fig:channels} show that the model cylinders fill the gaps of emission at the center of MM1, and that they also help to reproduce the velocity dispersion between compact sources. Our small model filaments are homogeneous and do not reproduce the clumpiness suggested by the data.

Something worth noting is that fixing the starting and ending point of the cylindrical flows, plus the above mentioned dynamical initial condition, automatically sets the line-of-sight arrangement of sources, allowing us to fully determine the 3D structure of the model cluster. 

\subsubsection{The entire W33A MM1 region}

Besides the localized features, the model also successfully reproduces the global features of W33A MM1. Figure 
\ref{fig:moments} shows that the line intensity is dominated by the north-south elongated emission coming from compact sources Main, MNE, S, and SNW, with secondary peaks at the positions of SE, E, and Ridge, and extended emission along the spiral-like filament and in the junctions between compact sources. The overall velocity field is also well reproduced: the brightest peak in Main is the most blueshifted, as well as the area going north-south from Main to S on the east side of S. The east (blueshifted) to west (redshifted) velocity gradient centered on S is also  reproduced. The jump to redshifted emission going from Main to MNE is also apparent, as well as the middle-velocity (green) valley between MNE-Main-S-SNW and E-SE, and redshifted emission at the positions of E and on the west side of SE. The overall velocity pattern of the spiral-like filament is also reproduced, as described in the previous section. The observed MM1 core is somewhat more extended than the model, probably due to extended emission not belonging to any compact source or filament. 
The velocity dispersion maps also match well. The highest velocity dispersion is localized around the most massive compact source Main, with high peaks around S. However, the model is short in the velocity dispersion around compact sources SE and Ridge. This extra observed velocity dispersion could be due to contamination of neighboring molecular lines within the velocity integration range for these two sources (see Fig. \ref{fig:spectra_sources}).  

\begin{figure*}
 \centering
 \includegraphics[width= 0.9\textwidth]{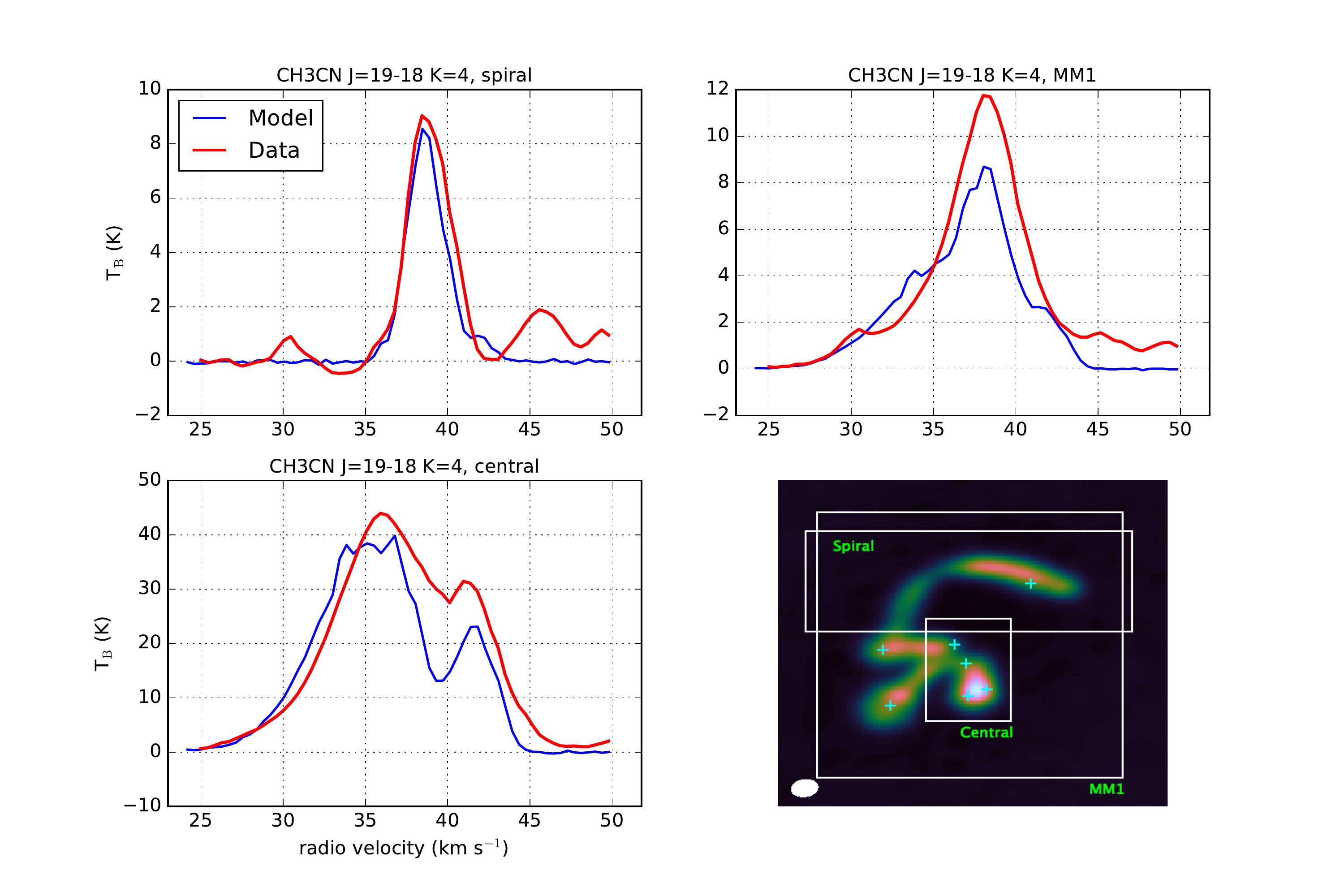} 
 \caption{CH$_3$CN $J=19-18$, $K=4$ line emission of model and ALMA data compared in large apertures of interest. The header on each panel indicates the zone of analysis. The bottom right panel is a snapshot of the cube in the velocity channel $v = 38.05$ km s$^{-1}$, where white squares show the areas over which the spectra have been averaged. The cyan markers indicate the center of the compact sources included in the model.}  
\label{fig:spectra_regions}
\end{figure*}

Figure \ref{fig:spectra_regions} shows a comparison of the model and observed spectra averaged in larger apertures containing the entire MM1 core and a region covering the brightest emission (Main+MNE+S+SNW), labeled as `central'. In the former, it is clear that the model reproduces the line centroid and width but is lacking about one third of the peak aperture-averaged brightness temperature, i.e., the missing extended emission mentioned above. The match of the model in the `central' part is better, although still some brightness from extended emission is missing. Appendix \ref{ap:chanmaps} shows the channel maps of both the model and ALMA data for further comparison. It is also clear from these that some extended emission is missing in the model, but that such emission could not be modelled with a core-scale, Ulrich-type or spherical envelope.

\subsection{(Sub)millimeter dust continuum} \label{continuum}

\begin{figure*}  
 \centering
 \includegraphics[width= 0.9\textwidth]{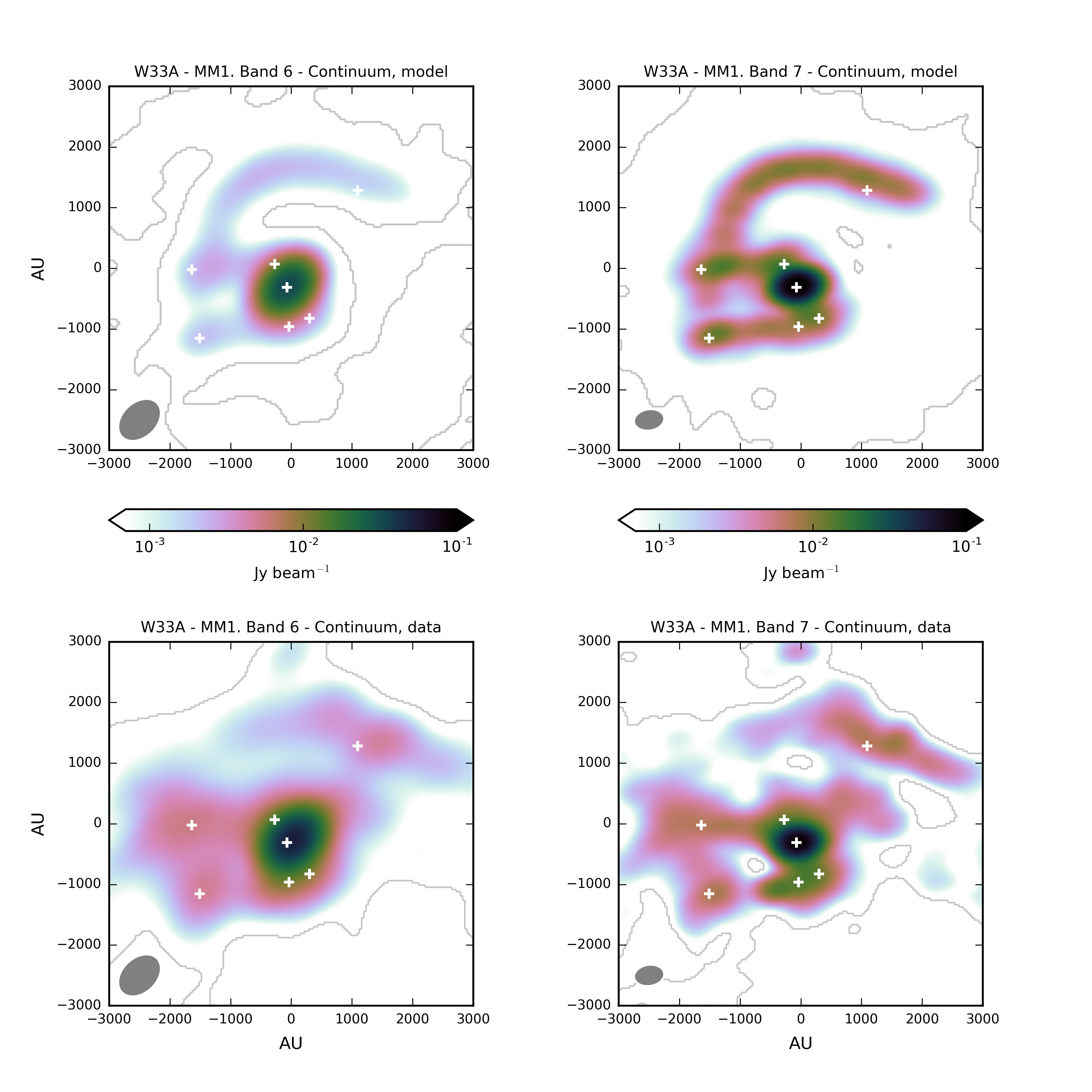} 
 \caption{220 GHz ({\it left} column) and 349 GHz ({\it right} column) continuum images of W33A MM1. {\it Top} panels correspond to the model and {\it bottom} panels to the observations. The compact sources included in the model are marked with crosses (+). The beam is shown in the lower left corner of each panel. The color bars are shared between panels of the same column.}  
\label{fig:continuum}
\end{figure*}

Model continuum maps of the thermal dust emission at 220.8 GHz (1.36 mm) and 349.3 GHz (0.86 mm) and the corresponding ALMA maps are shown together in Fig. \ref{fig:continuum}. We use an opacity power-law $\kappa = \kappa_0 (\nu/ \nu_0)^\beta$ with an opacity index $\beta = 1.7$, typical of the ISM, and a normalization $\kappa_0 = 0.5$ cm$^2$ g$^{-1}$ at 220 GHz as in \cite{GM10}. 
The global model was chosen to match well the 0.86 mm continuum and CH$_3$CN $J=19-18$ line, and then the resulting 1.36 mm flux is calculated. 

In Table \ref{table:3a} we list the peak and averaged continuum intensities over the same apertures used for the line analysis (Figures \ref{fig:spectra_sources} and \ref{fig:spectra_regions}). We avoid quoting fluxes\footnote{Throughout this paper we use the word `flux' to refer to a flux density, defined as a solid-angle integrated intensity.} 
for the following reasons:
i) The Band 7 and Band 6 beams are about- and larger than the $0.2\arcsec$ apertures that we use for the compact sources, respectively, and the sources are also of the order of this size. Thus, fluxes measured over these apertures do not exactly correspond to the correct source flux. ii) In some cases there is crowding between the compact sources and also with the filaments. Selecting larger apertures compared to the beam would help to solve the previous point i), but then the fluxes do not correspond anymore to those from individual objects. Selecting smaller apertures helps to isolate individual sources, but the situation of point i) gets worse.

For compact source Main, we found that a disc is needed to match the high and compact continuum intensities. Without a disc, a pure envelope only produces $\sim 10\%$ of the needed continuum emission, and its appearance is more extended than in the observations. This is a natural consequence of the envelope being less compact than the disc (see Fig. \ref{fig:disc_envelope}). Also, the compact appearance of the continuum and the line data suggests that the disc around Main should be small. Table \ref{table:1} shows that the estimated disc radius (152 au) is the smallest among all sources.


\setlength{\tabcolsep}{5pt} 

\begin{table*}
\centering
{\renewcommand{\arraystretch}{1.25}

\resizebox{\textwidth}{!}{
\begin{tabular}{@{} l *{10}{c}  @{} }
\toprule
\multicolumn{1}{c}{\thead{Region}} & \multicolumn{4}{c}{Model} & & \multicolumn{4}{c}{Data}  \\ 
\cline{2-5} \cline{7-10} & \thead{220 GHz \\ Mean [mJy/bm]} & \thead{220 GHz \\ Peak [mJy/bm]} & \thead{349 GHz \\ Mean [mJy/bm]} & \thead{349 GHz \\ Peak [mJy/bm]} & & \thead{220 GHz \\ Mean [mJy/bm]} & \thead{220 GHz \\ Peak [mJy/bm]} & \thead{349 GHz \\ Mean [mJy/bm]} & \thead{349 GHz \\ Peak [mJy/bm]} \\   \hline

Main & 26.2 & 33.6 & 67.3 & 119.7 & & 25.0 & 44.9 & 56.8 & 96.9 \\
MNE & 10.8 & 29.8 & 19.4 & 79.1 & & 18.6 & 49.4 & 18.1 & 65.8 \\
S & 5.7 & 15.0 & 9.0 & 15.0 & & 12.2 & 23.6 & 12.7 & 16.3 \\
SNW & 4.6 & 18.4 & 9.7 & 25.8 & & 11.1 & 30.7 & 11.5 & 24.4 \\
SE & 1.0 & 2.1 & 3.6 & 12.7 & & 2.0 & 1.8 & 3.6 & 7.9 \\
E & 2.1 & 2.9 & 8.5 & 14.5 & & 5.1 & 5.5 & 6.4 & 7.7 \\
Ridge & 1.4 & 1.9 & 6.6 & 10.3 & & 3.6 & 4.7 & 5.9 & 8.0 \\
Main-S & 17.9 & 32.9 & 33.4 & 110.2 & & 28.8 & 50.4 & 30.2 & 87.9 \\ 
Spiral & 0.5 & 3.3 & 1.6 & 10.9 & & 1.5 & 8.9 & 1.4 & 8.7 \\ 
Central-MM1 & 6.3 & 33.6 & 11.7 & 119.7 & & 11.4 & 44.9 & 12.3 & 96.9 \\ 
MM1 & 1.0 & 33.6 & 2.4 & 119.7 & & 2.2 & 44.9 & 2.3 & 96.9 \\ 

 \bottomrule
 \end{tabular} 
 
 }
 
\caption{Mean and peak continuum intensities of the modelled and observed regions. The known free-free contributions from Main and SE were extracted in Main, SE, Central-MM1 and MM1 for the data. The measurement apertures are the same as in Figures \ref{fig:spectra_sources} and \ref{fig:spectra_regions} except for SE, where we used a larger aperture to subtract adequately the free-free contribution. The absolute uncertainties are $\approx 10\%$ for both models and observations. 
}
\label{table:3a}
}
\end{table*}

A disc is also necessary to match the continuum emission of compact sources S, SE, and Ridge. For S, the mean intensity is dominated by the disc, with important contributions from the SE$\rightarrow$S filament and the close companion SNW. These external agents help to reproduce the horizontally elongated continuum emission in Band 7 around S (see Figure \ref{fig:continuum}). The mean intensity in SE has a significant contribution from the three 
filamentary flows coming/going from/to other sources. 
Source Ridge alters the appearance of the spiral-like filament, and a disc is needed to match the observational data. On the other hand, MNE and SNW are not required to host discs to reproduce their observed continuum. 

The mean intensity of the model spiral-like filament matches well the observations in both bands, but the real emission has inhomogeneities besides compact source Ridge that are not included in the model. An increased density toward Ridge would improve the match. 

The need for the filamentary flows joining pairs of compact sources is apparent in the continuum images. 
The bright, elongated features in the middle zones between compact sources  are well reproduced by the global model thanks to the inclusion of the cylindrical filaments. Similarly to the case of the line emission, the model lacks some extended emission that could arise from core emission not belonging to any of the compact sources or filaments. This extended emission, although morphologically noticeable in the real observations, amounts to only $22\%$ and $10\%$ flux on top of what the model respectively has at 1.3 mm and 0.8 mm, which is within the nominal $10 \%$ error in the observational flux determinations. 

Spectral indices were calculated for the large apertures shown in Figure \ref{fig:spectra_regions}. 
The model integrated fluxes for the spiral-like filament, the central-MM1 region, and the entire MM1 core, respectively at 1.3 and 0.8 mm are: 
$S_\mathrm{spiral,1.3mm}=30.0$ mJy,  
$S_\mathrm{spiral,0.8mm}=85.6$ mJy, 
$S_\mathrm{central,1.3mm}=51.1$ mJy, 
$S_\mathrm{central,0.8mm}=194.0$ mJy, 
$S_\mathrm{MM1,1.3mm}=102.4$ mJy, and
$S_\mathrm{MM1,0.8mm}=320.6$ mJy. 
The respective fluxes in the ALMA data are: 
$S_\mathrm{spiral,1.3mm}=34.0$ mJy,  
$S_\mathrm{spiral,0.8mm}=85.2$ mJy, 
$S_\mathrm{central,1.3mm}=71.1$ mJy, 
$S_\mathrm{central,0.8mm}=203.0$ mJy, 
$S_\mathrm{MM1,1.3mm}=125.0$ mJy, and
$S_\mathrm{MM1,0.8mm}=341.1$ mJy. 
The free-free contributions were subtracted from the observational data extrapolating the fluxes of the 7 mm sources in \cite{vdTMenten05} and using a free-free spectral index of 1 \citep[see also][]{Maud+17,GM10}. Thus, the obtained model dust spectral indices are:
$\alpha_\mathrm{spiral}=2.3$, $\alpha_\mathrm{central}=2.9$, and 
$\alpha_\mathrm{MM1}=2.5$. The observational spectral indices are: $\alpha_\mathrm{spiral}=2.0$, $\alpha_\mathrm{central}=2.3$, and 
$\alpha_\mathrm{MM1}=2.2$. The observational indices appear to be systematically lower than in the model, but taking into account the absolute uncertainties of about 10\% for both sets of images, 
the associated error in the spectral index calculation is $\pm$0.3, which makes the model and ALMA measurements consistent with each other.         

We note that the measured spectral indices in the model do not correspond to the $2+\beta$ that is often expected, and that is valid only under the Rayleigh-Jeans approximation and the optically-thin regime \citep[e.g.,][]{Maud13b}. Optical depth maps show that the model regions are optically thin on average except for the central parts of Main and SE, where the mean $\tau > 0.5$.  
Therefore, we interpret the low spectral indices as due to significant portions of the model being out of the Rayleigh-Jeans regime: $h\nu$ is in general less than $kT$, but not much less over large volumes. For example $h\nu/kT\approx0.1$ in the spiral-like filament. 
Although the central region of MM1 has an elevated optical depth, it is the closest to following the Rayleigh-Jeans limit, given that the temperatures there are also substantially higher.
It is possible that the origin of the low spectral indices in the real ALMA maps is the same, warning against readily interpreting low dust-emission spectral indices as a signature of grain growth when observations at frequencies larger than 300 GHz are used.  

\subsection{CH$_3$CN $J=19-18$, $K=8$} \label{sec:K8}

We now present a calculation of the CH$_3$CN $J=19-18$ $K=8$ line based on the previously described global model that matches the $K=4$ and continuum observations. For this line, the model emission is not meant to be a match to the data, but rather a check of how representative it is of the gas at scales smaller than the current observational angular resolution.  

\begin{figure*}  
 \centering
 \includegraphics[width= 0.9\textwidth]{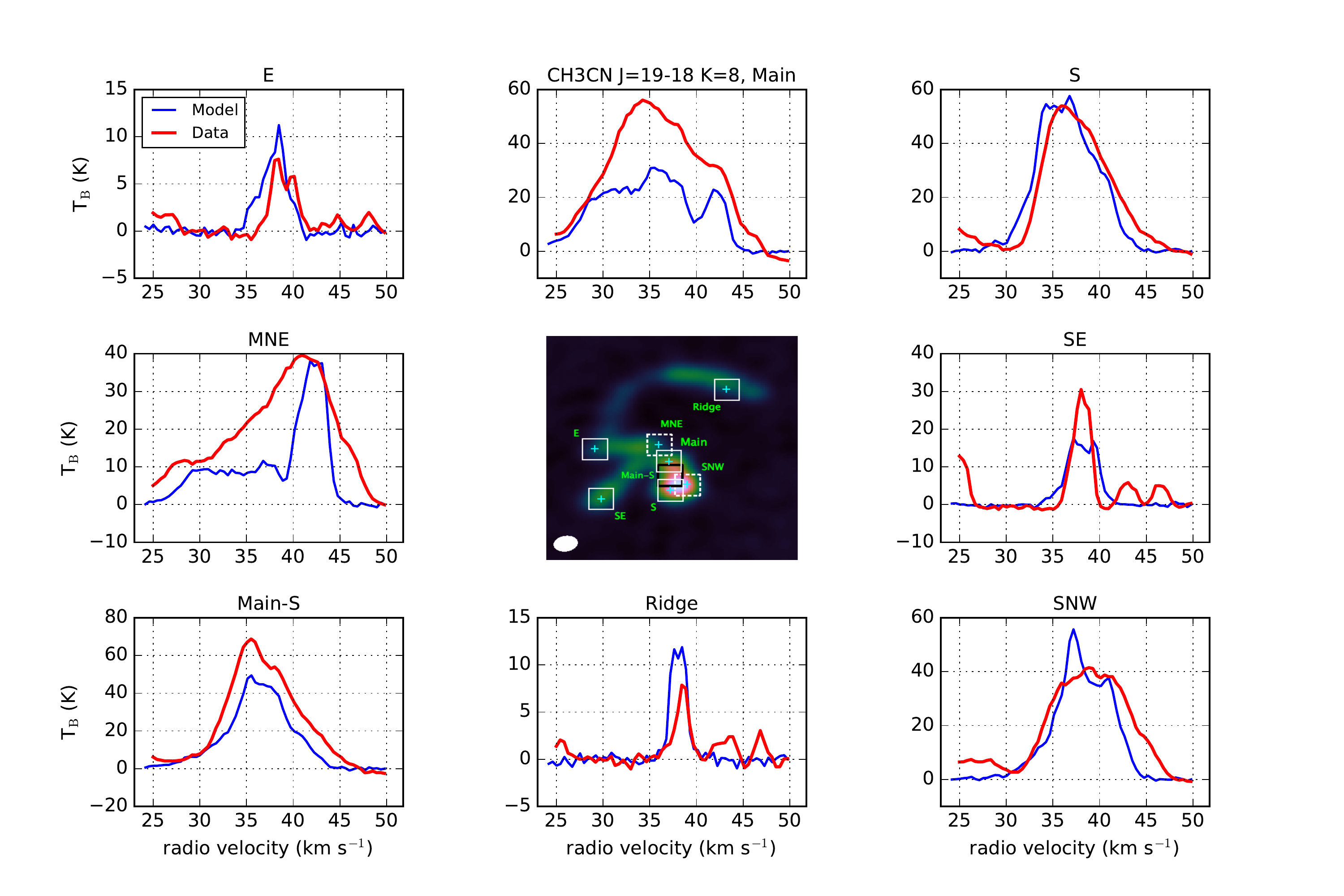} 
 \caption{Same as Figure \ref{fig:spectra_sources} for CH$_3$CN $J=19-18$, $K=8$.}  
\label{fig:spectra_sources_K8}
\end{figure*}

\begin{figure*}  
 \centering
 \includegraphics[width= 0.9\textwidth]{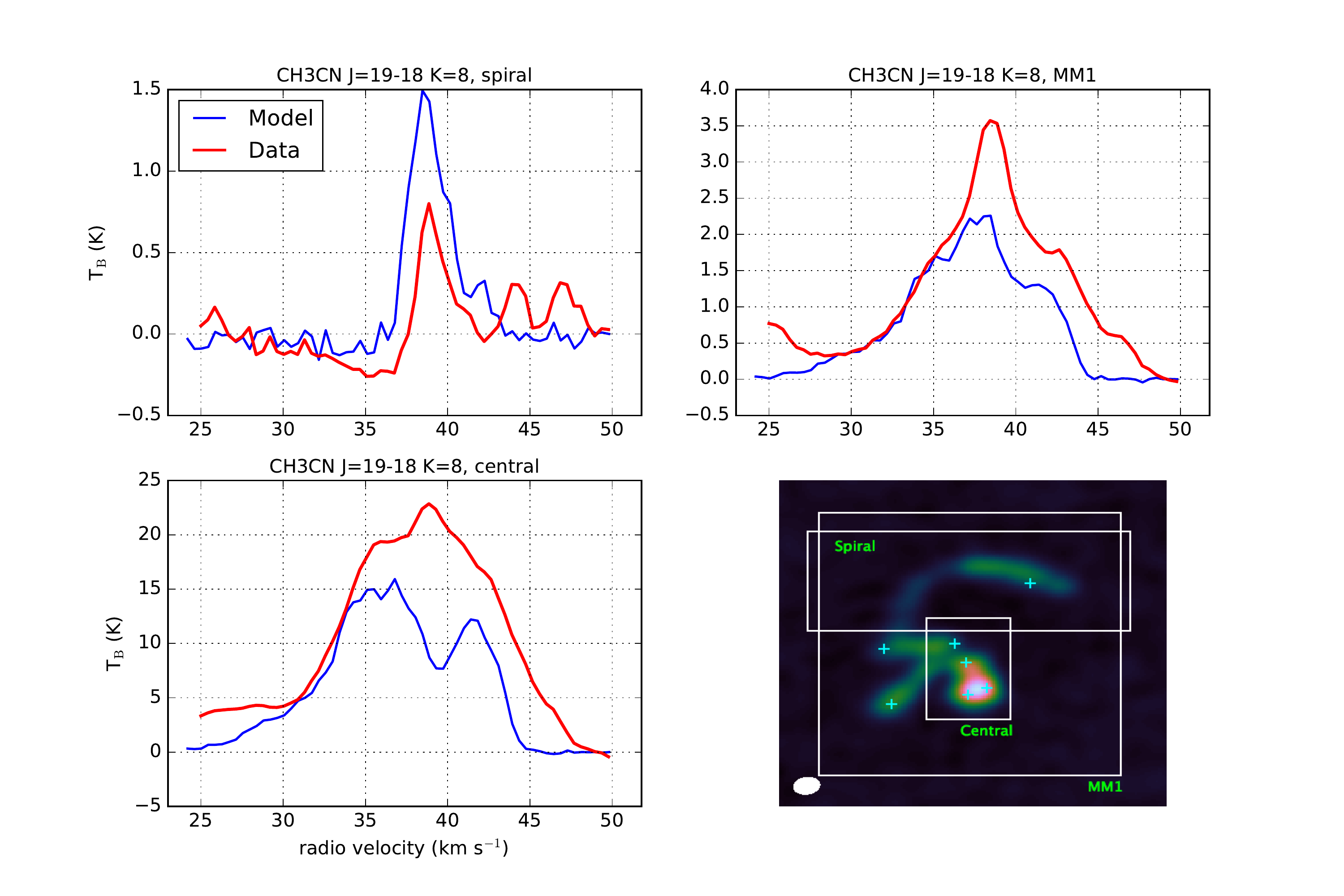} 
 \caption{Same as Figure \ref{fig:spectra_regions} for CH$_3$CN $J=19-18$, $K=8$.}  
\label{fig:spectra_regions_K8}
\end{figure*}

Figures \ref{fig:spectra_sources_K8} and \ref{fig:spectra_regions_K8} respectively show the spectra around the compact sources and on the same extended areas of interest as in the presentation of the $K=4$ model. The brightness match is reasonably good for all compact sources except for Main, and the line-width is only unmatched for MNE. The brighter observed line in Main suggests that the model should be warmer at radii $< 100$ au. The broader observed line in MNE could be due to contamination from Main. We expect to obtain ALMA long-baseline data in the near future to disentangle this crowded region and produce a more detailed model of the Main-MNE system. The match to the extended areas (Fig. \ref{fig:spectra_regions_K8}) is within a factor of $2$ in brightness for the large filament and the entire MM1, and better for the central region. In the latter, there is some missing {\bf emission} in the model at velocities close to the peak velocity of the entire MM1 core ($\approx 38$ km s$^{-1}$). This peak is well reproduced by the $K=4$ model (see Fig. \ref{fig:spectra_regions}).

\section{Discussion} \label{sec:discussion}

\subsection{An accretion filament feeding the fragmented high-mass core W33A MM1} \label{sec:feeding_filament} 

One of the main results of this study is to show, via 3D radiative transfer modelling, that the elongated structure north of the MM1 core is an accretion (feeding) flow. The observed kinematics and modelling show that the filament has both longitudinal and radial motions, and that it is filled with high-density ($\approx 8 \times 10^7$ cm$^{-3}$, see Table \ref{table:2}), warm ($\sim 200$ K) gas. The velocity dispersion within this spiral-like filament could only be reproduced if an extra, radial velocity component was added (see Section \ref{sec:acc_filaments-results}). Given the high density and the existence of fragmentation within the filament -- compact source Ridge --, infall is a plausible explanation to the observed extra velocity dispersion. 

The total gas mass in the model `feeding' filament is 0.41 $M_\odot$, and its average flow rate is $4.75\times10^{-5}~M_\odot$ yr$^{-1}$. It is possible that these are lower limits, since less dense, colder gas in the flow would not emit significantly in the CH$_3$CN lines. For the modelled mass inflow rate, the depletion time of the filament gas is $\sim 8.3\times10^3$ yr, quite short compared with the few $\times 10^5$ yr expected for massive star formation \citep{ZY07}. At first sight, this suggests that the `feeding filament' could be a transient structure, unless replenishment from larger scales occurs. Several studies have shown evidence for the continuity of molecular-gas flows from scales of $\sim 10$ pc down to $<0.05$ pc \citep[e.g.,][]{GM09,Schneider10,Liu12,Nakamura12,Peretto13}. 
For the case of W33A, \cite{GM10} reported that MM1 appears to be connected to MM2 by an extension of gas 
$\sim 12 000$ au long in the northeast-southwest direction, similar to the orientation of the spiral-like filament. This possible larger-scale extension of the filament, however, is not modelled in this paper since it does not emit significantly in the observed CH$_3$CN transitions. Further evidence for replenishment in W33A comes from the pc-scale filaments seen in NH$_3$ emission, which converge in position-position-velocity space at the position of MM1 \citep{GM10}. 

The rate at which the model spiral-like filament provides mass to MM1 is an order of magnitude below the combined protostellar accretion rate of the model compact sources $\sim 4.6\times10^{-4}~M_\odot$ yr$^{-1}$  (see Table \ref{table:1}), and dominated by the accretion onto Main. We consider two possible interpretations for this: the mismatch between the protostellar and core accretion rates could mean that the former will be significantly lower within a few $\times 10^4$ yr, after the gas reservoir in the modelled filaments is depleted. This is consistent with the onset of ionization in source Main, which hosts a tiny hypercompact HII region with an estimated size $< 100$ au \citep{vdTMenten05}. On the other hand, it is possible that gas accretion will continue for longer timescales if there is the aforementioned replenishment from larger scales and/or our mass estimates for the intra-core filaments are lower limits because some gas does not emit in the modelled CH$_3$CN lines.

Spiral-like structures like the filament feeding MM1 have been observed both at smaller and larger scales, from low-mass protoplanetary discs/envelopes \citep[$10^2$ au,][]{Perez16,Yen17}, to luminous, cluster-forming clumps 
\citep[$10^5$ au,][]{Wright14,Liu15}. 
Such spiral structures feeding material to nascent stellar systems form as a natural consequence of gravitational fragmentation in (radiation) hydrodynamical simulations \citep[e.g.,][]{Bate11,Vorobyov13}. Our observations and analytical model look  similar to the simulations of massive star formation presented by \cite{Krumholz07}, who calculated specific predictions for images taken with ALMA in CH$_3$CN transitions. The synthetic images from those simulations show spiral filaments with typical length scales of a few thousand au, feeding a central object that reaches a stellar mass of $8~M_\odot$ and a few lower-mass companions. Their total gas mass within $\sim 1000$ au of the central core reaches about $5~M_\odot$. These characteristics are similar to those of our analytical model, although in the \cite{Krumholz07} simulations such structures arise in the context of a massive fragmenting disc, whereas our model is ad-hoc.

\subsection{Accretion filaments joining pairs of protostars} 

Another main result of this study is the proposed existence of gas flows between pairs of compact sources in the cluster-forming environment of W33A MM1. The ALMA continuum observations, especially the higher angular resolution 0.8-mm image, clearly shows elongations joining some of the compact sources (see Fig. \ref{fig:continuum}). The two most notable features are between sources E and Main/MNE, and between SE and S. 
Table \ref{table:2} shows that these two are the most massive among the small filaments, with $\sim 0.04~M_\odot$ each. 
Our interpretation also explains the observed CH$_3$CN intensity, velocity field, and velocity dispersion in these inter-source regions (see Fig. \ref{fig:moments}). The existence of these features is a robust result, however, we note that our implementation of their kinematics and morphology is just a first-order approximation. Also, the existence of the two `diagonal' filaments is not without doubt, but their implementation helped to better reproduce a few fine details seen in the data (Section \ref{sec:acc_filaments-results}).

The accretion rate across the E $\rightarrow$ MNE filament is half of the feeding rate of the larger, spiral-like filament, suggesting that most of the gas processed by the latter ends up in the Main/MNE system. This finding is consistent with models of star (cluster) formation that emphasize the need for replenishment of gas from cloud to clump to core to protostellar scales \citep[e.g.,][]{Bonnell04,Smith09,BP15,VS17}.  

\subsection{The protostellar objects}

One of the unexpected findings of our modelling is that the mass of the object at the center of compact source Main is lower ($M_{\star,Main}=7~M_\odot$) than previous estimates. Retrospectively, this is not surprising, since the increment in angular resolution ($\times9$ in beam area) has unveiled that what we thought was a system with two sources (Main and SE) has many more components. The total dynamical mass of the sources in the clustered area (Main+MNE+S+SNW) amounts to $11.43~M_\odot$, of which $11~M_\odot$ are stellar and $0.43~M_\odot$ are gas. This is in good agreement with \cite{GM10}, who estimated a stellar mass $\sim 10~M_\odot$. The gas mass that can be inferred from their peak intensity in Main, which at their resolution roughly corresponds to the clustered area in the ALMA data, and after correcting for free-free emission and re-scaling for the new distance, is $\sim 0.35~M_\odot$, also similar to our model result. 

With a mass $M_{\star,Main}=7~M_\odot$, Main is at the border of what usually are called intermediate- and high-mass stars. The modelling of the gas flows suggests that it could accrete a large fraction of the $\sim 1~M_\odot$ gas reservoirs seen in the ALMA data (see Tables \ref{table:1} and \ref{table:2}). Also, Main could accrete more gas if the aforementioned replenishment occurs, although it is uncertain.
In the following, we check if the modelled YSO with a stellar mass $M_{\star,Main}=7~M_\odot$ is still consistent with the high luminosity of W33A.  
We note that the (proto)stellar luminosity is not a free parameter in our modelling, since the temperature structure of the discs and envelopes are manually set following the prescriptions of Section \ref{sec:physmodel}. However, it can be checked what is the resulting luminosity from the selected parameters (see Table \ref{table:1}).  Taking the accretion luminosity as $L_{acc}=(G M_\star \dot{M}) / R_\star$, we obtain for Main that $L_{acc} = 2.9\times10^3~L_\odot$.
The accretion rate sets the density structure, the stellar mass sets the kinematics of the gas around it, and the stellar radius has a small effect on the model because the normalization of the disc scale height is proportional to it (see Section \ref{sec:physmodel}).
The complementary stellar luminosity would be $L_\star=4\pi R_\star^2 \sigma_\mathrm{SB} T_{eff}^4$. A (proto)stellar object bloated by accretion as the one we consider will have an effective temperature $T_{eff}$ lower than the corresponding ZAMS star of the same mass \citep{Hosokawa10}. Considering $T_{eff} \approx 10^4$ K, the corresponding stellar luminosity is $L_\star \approx 7.9 \times 10^3~L_\odot$. Thus, our model gives a total luminosity for Main $L_\mathrm{Main} \approx 1.1 \times 10^4~L_\odot$. The luminosity of W33A on scales of a few arcmin was considered to be $L_\mathrm{W33A} \sim 1\times10^5~L_\odot$ \citep{Stier84} for an assumed distance of 3.8 kpc, but the new parallax distance of 2.4 kpc \citep{Immer13} lowers this estimation to $L_\mathrm{W33A} \approx 4 \times 10^4~L_\odot$. \cite{Lin16} published SED fitting of molecular clouds at 10 arcsec resolution from combining ground-based bolometric and {\it Herschel} data. We requested their W33 images and measured the bolometric luminosity within one beam area around the W33A peak, which roughly corresponds to the area of W33A MM1+MM2. We obtain $L_\mathrm{MM1+MM2} \approx 1.8\times10^4~L_\odot$. Therefore, the luminosity that our model implies for Main and the observed luminosity at the smallest scales in which it can be measured are consistent with each other.

Considering the other sources, we find that some of them do require discs (S, SE, and Ridge), but some others do not (MNE, SNW, E), to simultaneously match their spectra and continuum (Section \ref{sec:results}). The sources that require discs host the most massive or more evolved protostellar objects, which could either mean that the least massive sources are younger, possibly low-mass class 0 YSOs where disc formation is still occurring \citep[e.g.,][]{Li17}, or that our observations do not have the sensitivity and resolution to properly constrain the discs that could be embedded within these fainter sources. The disc+envelope to (proto)stellar mass ratio is 0.056 for Main and in the range 0.015 to 0.044 for the rest of the sources (E is a pure envelope). These values are not accurately constrained, but illustrate that most of the compact sources are stellar rather than gas dominated, although a fraction of the extended emission in the core could be resolved out by the interferometer. 

\subsection{A high-mass stellar association in the making}

Our modelling gives a distribution of stellar masses within the MM1 core with one massive protostellar object (Main), one of intermediate mass (S), and four of low mass (SE, MNE, SNW, Ridge), as well as one pre-stellar core (E) without a central object. This is loosely reminiscent of a stellar initial mass function \citep[IMF,][]{Salpeter55}, although we emphasize that our number of protostellar sources is small. This result suggests that several physical processes within the MM1 core, such as further fragmentation into the individual compact sources, as well as gas flows between them and from their environment \citep[e.g.,][]{Rosen16,Peters10}, could be relevant for the setting of the IMF. 
These factors could add in a complex manner to processes that operate at larger scales \citep[e.g.,][]{Offner14,Oey11}, like those behind determining the mass distribution of cores such as MM1, 
which also appears to be a decreasing function with object mass 
\citep[e.g.,][]{Alves07}.

The number of compact sources in our model translates into an average (proto)stellar density of $1.8\times10^5$ pc$^{-3}$, which is still lower than the $10^6$ to $10^8$ pc$^{-3}$ required for (proto)stellar collisions to work as a significant agent in shaping the nascent stellar association \citep{ZY07}. Since we detect objects down to a fraction of a solar mass and sources of lower mass are not relevant for the required gravitational focusing, we argue that we can rule out stellar collisions in this particular region. 

Close encounters, however, could still be relevant. It can be seen from Table \ref{table:1} that the inferred disc radii are  from 150 to 350 au, in good agreement with the observed sizes of low-mass protoplanetary discs \citep[e.g.,][]{Andrews15}. This is expected for the low-mass proto(stars) in our model but is not obvious for Main. One possibility is that the disc of Main has been truncated due to interactions with the implied nearby (474 au) source MNE \citep[e.g.,][]{Vincke15}, or even that MNE is the result of the fragmentation of the disc around Main \citep[e.g.,][]{Vorobyov13}. 

The MM1 forming association appears to be virialized. Taking into account the entire model mass and a radius of 3500 au, the 1D escape velocity of the MM1 core is $\approx 1.3$ km s$^{-1}$. Similarly, the 1D rms velocity dispersion of the systemic velocities of the compact sources with respect to the systemic velocity of the entire gas core ($\approx 38$ km s$^{-1}$) is $\approx 1.5$ km s$^{-1}$. It could be expected that after star formation is shut off and the remaining core gas is removed, the resulting stellar association will be super-virial, i.e., it will have a velocity dispersion larger than the equilibrium one and will be dissolved within a few Myr \citep[e.g.,][]{GoodwinBastian06}. 

\section{Conclusions} \label{sec:conclusions}

We have made a multiple-component analytical model of the complex massive star formation region W33A MM1, and performed radiative-transfer calculations using LIME to predict its observational appearance and compare it to ALMA images at $\approx0.2$ arcsec resolution. The model was tailored to match CH$_3$CN lines and dust continuum emission from dense and warm gas. Our main conclusions are as follows: 
\begin{itemize}
\item The MM1 core is fragmented into six compact sources within a 1000 au radius, plus another compact source within the `feeding' spiral-like filament. Some of these sources require the presence of a disc within an envelope to simultaneously match the high continuum and line intensities, whereas some others can be modelled as pure envelopes.
\item Compared to previous estimates, we obtain lower masses ($M_\star \approx 7~M_\odot$, $M_\mathrm{disc+envelope} \approx 0.4~M_\odot$) and a smaller disc size ($R_\mathrm{d} \sim 150$ au) for the most luminous (proto)star in the region, known as Main ($L_\mathrm{Main} \sim 1.1\times10^4~L_\odot$). This is a consequence of the high-level of fragmentation found within the core. The total dynamical (stellar+gas) mass of our model is consistent with previous estimations. 
\item The spiral-like filament converging to MM1 from the northwest can be convincingly interpreted as an accretion flow feeding the nascent stellar association. The kinematics of this $\sim 10^4$ au length filament is consistent with a parabolic trajectory with focus at the center of mass of the MM1 cluster. The filament itself is fragmenting and appears to have a radial infall velocity component.

\item Small filamentary flows of $\sim 1000$ au length between pairs of (proto)stellar sources are proposed to exist. The most prominent one, from source E to the massive Main/MNE system, appears to hoard most of the gas flow rate coming from the larger, spiral-like filament that feeds the entire MM1 core. Gas replenishment from clump to core to protostellar scales appears to be key.
\item The forming stellar association seems to be virialized and may become super-virial if the remaining gas is removed, favouring the evaporation of the newly formed stars into the field. The distribution of (proto)stellar masses is such that there are several low-mass objects per high-mass star.
\end{itemize}

\section*{Acknowledgements}
The authors thank the referee for their useful reports.

AI thanks the support from \textit{Facultad de Ciencias Exactas y Naturales}-UdeA, \textit{Relaciones Internacionales}-UdeA, IRyA-UNAM and the The Leiden/ESA Astrophysics Program for Summer Students 2016. RGM acknowledges support from UNAM-PAPIIT program IA102817 and IRyA-UNAM.

\section*{Additional software}
In addition to the software referenced throughout the article, we used specific Python packages to achieve the modelling: Numpy \citep{Numpy}, Matplotlib \citep{Matplotlib}, Astropy \citep{Astropy}, IPython \citep{IPython} and Pandas \citep{Pandas}.

\bibliography{refs_W33A-MM1}

\begin{thebibliography}{}
\makeatletter
\relax
\def\mn@urlcharsother{\let\do\@makeother \do\$\do\&\do\#\do\^\do\_\do\%\do\~}
\def\mn@doi{\begingroup\mn@urlcharsother \@ifnextchar [ {\mn@doi@}
  {\mn@doi@[]}}
\def\mn@doi@[#1]#2{\def\@tempa{#1}\ifx\@tempa\@empty \href
  {http://dx.doi.org/#2} {doi:#2}\else \href {http://dx.doi.org/#2} {#1}\fi
  \endgroup}
\def\mn@eprint#1#2{\mn@eprint@#1:#2::\@nil}
\def\mn@eprint@arXiv#1{\href {http://arxiv.org/abs/#1} {{\tt arXiv:#1}}}
\def\mn@eprint@dblp#1{\href {http://dblp.uni-trier.de/rec/bibtex/#1.xml}
  {dblp:#1}}
\def\mn@eprint@#1:#2:#3:#4\@nil{\def\@tempa {#1}\def\@tempb {#2}\def\@tempc
  {#3}\ifx \@tempc \@empty \let \@tempc \@tempb \let \@tempb \@tempa \fi \ifx
  \@tempb \@empty \def\@tempb {arXiv}\fi \@ifundefined
  {mn@eprint@\@tempb}{\@tempb:\@tempc}{\expandafter \expandafter \csname
  mn@eprint@\@tempb\endcsname \expandafter{\@tempc}}}

\bibitem[\protect\citeauthoryear{{Alves}, {Lombardi}  \& {Lada}}{{Alves}
  et~al.}{2007}]{Alves07}
{Alves} J.,  {Lombardi} M.,   {Lada} C.~J.,  2007, \mn@doi [\aap]
  {10.1051/0004-6361:20066389}, \href
  {http://adsabs.harvard.edu/abs/2007A%26A...462L..17A} {462, L17}

\bibitem[\protect\citeauthoryear{{Andrews}}{{Andrews}}{2015}]{Andrews15}
{Andrews} S.~M.,  2015, \mn@doi [\pasp] {10.1086/683178}, \href
  {http://adsabs.harvard.edu/abs/2015PASP..127..961A} {127, 961}

\bibitem[\protect\citeauthoryear{{Araya}, {Hofner}, {Kurtz}, {Bronfman}  \&
  {DeDeo}}{{Araya} et~al.}{2005}]{Araya05}
{Araya} E.,  {Hofner} P.,  {Kurtz} S.,  {Bronfman} L.,   {DeDeo} S.,  2005,
  \mn@doi [\apjs] {10.1086/427187}, \href
  {http://adsabs.harvard.edu/abs/2005ApJS..157..279A} {157, 279}

\bibitem[\protect\citeauthoryear{{Astropy Collaboration} et~al.,}{{Astropy
  Collaboration} et~al.}{2013}]{Astropy}
{Astropy Collaboration} et~al., 2013, \mn@doi [\aap]
  {10.1051/0004-6361/201322068}, \href
  {http://adsabs.harvard.edu/abs/2013A%26A...558A..33A} {558, A33}

\bibitem[\protect\citeauthoryear{{Ballesteros-Paredes}, {Hartmann},
  {P{\'e}rez-Goytia}  \& {Kuznetsova}}{{Ballesteros-Paredes}
  et~al.}{2015}]{BP15}
{Ballesteros-Paredes} J.,  {Hartmann} L.~W.,  {P{\'e}rez-Goytia} N.,
  {Kuznetsova} A.,  2015, \mn@doi [\mnras] {10.1093/mnras/stv1285}, \href
  {http://adsabs.harvard.edu/abs/2015MNRAS.452..566B} {452, 566}

\bibitem[\protect\citeauthoryear{{Bate}}{{Bate}}{2011}]{Bate11}
{Bate} M.~R.,  2011, \mn@doi [\mnras] {10.1111/j.1365-2966.2011.19386.x}, \href
  {http://adsabs.harvard.edu/abs/2011MNRAS.417.2036B} {417, 2036}

\bibitem[\protect\citeauthoryear{{Beltr{\'a}n} \& {de Wit}}{{Beltr{\'a}n} \&
  {de Wit}}{2016}]{BeltranDeWit16}
{Beltr{\'a}n} M.~T.,  {de Wit} W.~J.,  2016, \mn@doi [\aapr]
  {10.1007/s00159-015-0089-z}, \href
  {http://adsabs.harvard.edu/abs/2016A%26ARv..24....6B} {24, 6}

\bibitem[\protect\citeauthoryear{{Bergin} \& {Tafalla}}{{Bergin} \&
  {Tafalla}}{2007}]{BerginTafalla07}
{Bergin} E.~A.,  {Tafalla} M.,  2007, \mn@doi [\araa]
  {10.1146/annurev.astro.45.071206.100404}, \href
  {http://adsabs.harvard.edu/abs/2007ARA%26A..45..339B} {45, 339}

\bibitem[\protect\citeauthoryear{{Beuther}, {Walsh}, {Johnston}, {Henning},
  {Kuiper}, {Longmore}  \& {Walmsley}}{{Beuther} et~al.}{2017}]{Beuther17}
{Beuther} H.,  {Walsh} A.~J.,  {Johnston} K.~G.,  {Henning} T.,  {Kuiper} R.,
  {Longmore} S.~N.,   {Walmsley} C.~M.,  2017, \mn@doi [\aap]
  {10.1051/0004-6361/201630126}, \href
  {http://adsabs.harvard.edu/abs/2017A%26A...603A..10B} {603, A10}

\bibitem[\protect\citeauthoryear{{Bonnell}, {Vine}  \& {Bate}}{{Bonnell}
  et~al.}{2004}]{Bonnell04}
{Bonnell} I.~A.,  {Vine} S.~G.,   {Bate} M.~R.,  2004, \mn@doi [\mnras]
  {10.1111/j.1365-2966.2004.07543.x}, \href
  {http://adsabs.harvard.edu/abs/2004MNRAS.349..735B} {349, 735}

\bibitem[\protect\citeauthoryear{{Brinch} \& {Hogerheijde}}{{Brinch} \&
  {Hogerheijde}}{2010}]{Brinch+10}
{Brinch} C.,  {Hogerheijde} M.~R.,  2010, \mn@doi [\aap]
  {10.1051/0004-6361/201015333}, \href
  {http://adsabs.harvard.edu/abs/2010A%26A...523A..25B} {523, A25}

\bibitem[\protect\citeauthoryear{{Bunn}, {Hoare}  \& {Drew}}{{Bunn}
  et~al.}{1995}]{Bunn95}
{Bunn} J.~C.,  {Hoare} M.~G.,   {Drew} J.~E.,  1995, \mn@doi [\mnras]
  {10.1093/mnras/272.2.346}, \href
  {http://adsabs.harvard.edu/abs/1995MNRAS.272..346B} {272, 346}

\bibitem[\protect\citeauthoryear{{Carrasco-Gonz{\'a}lez}
  et~al.,}{{Carrasco-Gonz{\'a}lez} et~al.}{2012}]{CG12}
{Carrasco-Gonz{\'a}lez} C.,  et~al., 2012, \mn@doi [\apjl]
  {10.1088/2041-8205/752/2/L29}, \href
  {http://adsabs.harvard.edu/abs/2012ApJ...752L..29C} {752, L29}

\bibitem[\protect\citeauthoryear{{Cesaroni}, {Felli}, {Jenness}, {Neri},
  {Olmi}, {Robberto}, {Testi}  \& {Walmsley}}{{Cesaroni}
  et~al.}{1999}]{Cesaroni99}
{Cesaroni} R.,  {Felli} M.,  {Jenness} T.,  {Neri} R.,  {Olmi} L.,  {Robberto}
  M.,  {Testi} L.,   {Walmsley} C.~M.,  1999, \aap, \href
  {http://adsabs.harvard.edu/abs/1999A%26A...345..949C} {345, 949}

\bibitem[\protect\citeauthoryear{{Cesaroni} et~al.,}{{Cesaroni}
  et~al.}{2017}]{Cesaroni17}
{Cesaroni} R.,  et~al., 2017, \mn@doi [\aap] {10.1051/0004-6361/201630184},
  \href {http://adsabs.harvard.edu/abs/2017A%26A...602A..59C} {602, A59}

\bibitem[\protect\citeauthoryear{{Cummins}, {Green}, {Thaddeus}  \&
  {Linke}}{{Cummins} et~al.}{1983}]{Cummins83}
{Cummins} S.~E.,  {Green} S.,  {Thaddeus} P.,   {Linke} R.~A.,  1983, \mn@doi
  [\apj] {10.1086/160782}, \href
  {http://adsabs.harvard.edu/abs/1983ApJ...266..331C} {266, 331}

\bibitem[\protect\citeauthoryear{{Davies}, {Lumsden}, {Hoare}, {Oudmaijer}  \&
  {de Wit}}{{Davies} et~al.}{2010}]{Davies10}
{Davies} B.,  {Lumsden} S.~L.,  {Hoare} M.~G.,  {Oudmaijer} R.~D.,   {de Wit}
  W.-J.,  2010, \mn@doi [\mnras] {10.1111/j.1365-2966.2009.16077.x}, \href
  {http://adsabs.harvard.edu/abs/2010MNRAS.402.1504D} {402, 1504}

\bibitem[\protect\citeauthoryear{{Evans}}{{Evans}}{1999}]{Evans99}
{Evans} II N.~J.,  1999, \mn@doi [\araa] {10.1146/annurev.astro.37.1.311},
  \href {http://adsabs.harvard.edu/abs/1999ARA%26A..37..311E} {37, 311}

\bibitem[\protect\citeauthoryear{{Galv{\'a}n-Madrid}, {Keto}, {Zhang}, {Kurtz},
  {Rodr{\'{\i}}guez}  \& {Ho}}{{Galv{\'a}n-Madrid} et~al.}{2009}]{GM09}
{Galv{\'a}n-Madrid} R.,  {Keto} E.,  {Zhang} Q.,  {Kurtz} S.,
  {Rodr{\'{\i}}guez} L.~F.,   {Ho} P.~T.~P.,  2009, \mn@doi [\apj]
  {10.1088/0004-637X/706/2/1036}, \href
  {http://adsabs.harvard.edu/abs/2009ApJ...706.1036G} {706, 1036}

\bibitem[\protect\citeauthoryear{{Galv{\'a}n-Madrid}, {Zhang}, {Keto}, {Ho},
  {Zapata}, {Rodr{\'{\i}}guez}, {Pineda}  \&
  {V{\'a}zquez-Semadeni}}{{Galv{\'a}n-Madrid} et~al.}{2010}]{GM10}
{Galv{\'a}n-Madrid} R.,  {Zhang} Q.,  {Keto} E.,  {Ho} P.~T.~P.,  {Zapata}
  L.~A.,  {Rodr{\'{\i}}guez} L.~F.,  {Pineda} J.~E.,   {V{\'a}zquez-Semadeni}
  E.,  2010, \mn@doi [\apj] {10.1088/0004-637X/725/1/17}, \href
  {http://adsabs.harvard.edu/abs/2010ApJ...725...17G} {725, 17}

\bibitem[\protect\citeauthoryear{{Ginsburg} et~al.,}{{Ginsburg}
  et~al.}{2017}]{Ginsburg17}
{Ginsburg} A.,  et~al., 2017, \mn@doi [\apj] {10.3847/1538-4357/aa6bfa}, \href
  {http://adsabs.harvard.edu/abs/2017ApJ...842...92G} {842, 92}

\bibitem[\protect\citeauthoryear{{Girart} et~al.,}{{Girart}
  et~al.}{2018}]{Girart18}
{Girart} J.~M.,  et~al., 2018, \mn@doi [\apjl] {10.3847/2041-8213/aab76b},
  \href {http://adsabs.harvard.edu/abs/2018ApJ...856L..27G} {856, L27}

\bibitem[\protect\citeauthoryear{{Goodwin} \& {Bastian}}{{Goodwin} \&
  {Bastian}}{2006}]{GoodwinBastian06}
{Goodwin} S.~P.,  {Bastian} N.,  2006, \mn@doi [\mnras]
  {10.1111/j.1365-2966.2006.11078.x}, \href
  {http://adsabs.harvard.edu/abs/2006MNRAS.373..752G} {373, 752}

\bibitem[\protect\citeauthoryear{{Hosokawa}, {Yorke}  \& {Omukai}}{{Hosokawa}
  et~al.}{2010}]{Hosokawa10}
{Hosokawa} T.,  {Yorke} H.~W.,   {Omukai} K.,  2010, \mn@doi [\apj]
  {10.1088/0004-637X/721/1/478}, \href
  {http://adsabs.harvard.edu/abs/2010ApJ...721..478H} {721, 478}

\bibitem[\protect\citeauthoryear{Hunter}{Hunter}{2007}]{Matplotlib}
Hunter J.~D.,  2007, \mn@doi [Computing In Science \& Engineering]
  {10.1109/MCSE.2007.55}, 9, 90

\bibitem[\protect\citeauthoryear{{Hunter} et~al.,}{{Hunter}
  et~al.}{2017}]{Hunter17}
{Hunter} T.~R.,  et~al., 2017, \mn@doi [\apjl] {10.3847/2041-8213/aa5d0e},
  \href {http://adsabs.harvard.edu/abs/2017ApJ...837L..29H} {837, L29}

\bibitem[\protect\citeauthoryear{{Immer}, {Reid}, {Menten}, {Brunthaler}  \&
  {Dame}}{{Immer} et~al.}{2013}]{Immer13}
{Immer} K.,  {Reid} M.~J.,  {Menten} K.~M.,  {Brunthaler} A.,   {Dame} T.~M.,
  2013, \mn@doi [\aap] {10.1051/0004-6361/201220793}, \href
  {http://adsabs.harvard.edu/abs/2013A%26A...553A.117I} {553, A117}

\bibitem[\protect\citeauthoryear{{Immer}, {Galv{\'a}n-Madrid}, {K{\"o}nig},
  {Liu}  \& {Menten}}{{Immer} et~al.}{2014}]{Immer14}
{Immer} K.,  {Galv{\'a}n-Madrid} R.,  {K{\"o}nig} C.,  {Liu} H.~B.,   {Menten}
  K.~M.,  2014, \mn@doi [\aap] {10.1051/0004-6361/201423780}, \href
  {http://adsabs.harvard.edu/abs/2014A%26A...572A..63I} {572, A63}

\bibitem[\protect\citeauthoryear{{Johnston} et~al.,}{{Johnston}
  et~al.}{2015}]{Johnston15}
{Johnston} K.~G.,  et~al., 2015, \mn@doi [\apjl] {10.1088/2041-8205/813/1/L19},
  \href {http://adsabs.harvard.edu/abs/2015ApJ...813L..19J} {813, L19}

\bibitem[\protect\citeauthoryear{{Keto}}{{Keto}}{2007}]{Keto07}
{Keto} E.,  2007, \mn@doi [\apj] {10.1086/520320}, \href
  {http://adsabs.harvard.edu/abs/2007ApJ...666..976K} {666, 976}

\bibitem[\protect\citeauthoryear{{Keto} \& {Rybicki}}{{Keto} \&
  {Rybicki}}{2010}]{KetoRyb10}
{Keto} E.,  {Rybicki} G.,  2010, \mn@doi [\apj] {10.1088/0004-637X/716/2/1315},
  \href {http://adsabs.harvard.edu/abs/2010ApJ...716.1315K} {716, 1315}

\bibitem[\protect\citeauthoryear{{Keto} \& {Zhang}}{{Keto} \&
  {Zhang}}{2010}]{KetoZhang10}
{Keto} E.,  {Zhang} Q.,  2010, \mn@doi [\mnras]
  {10.1111/j.1365-2966.2010.16672.x}, \href
  {http://adsabs.harvard.edu/abs/2010MNRAS.406..102K} {406, 102}

\bibitem[\protect\citeauthoryear{{Keto}, {Caselli}  \& {Rawlings}}{{Keto}
  et~al.}{2015}]{Keto15}
{Keto} E.,  {Caselli} P.,   {Rawlings} J.,  2015, \mn@doi [\mnras]
  {10.1093/mnras/stu2247}, \href
  {http://adsabs.harvard.edu/abs/2015MNRAS.446.3731K} {446, 3731}

\bibitem[\protect\citeauthoryear{{Krumholz}, {Klein}  \& {McKee}}{{Krumholz}
  et~al.}{2007}]{Krumholz07}
{Krumholz} M.~R.,  {Klein} R.~I.,   {McKee} C.~F.,  2007, \mn@doi [\apj]
  {10.1086/519305}, \href {http://adsabs.harvard.edu/abs/2007ApJ...665..478K}
  {665, 478}

\bibitem[\protect\citeauthoryear{{Lada} \& {Lada}}{{Lada} \&
  {Lada}}{2003}]{LadaLada03}
{Lada} C.~J.,  {Lada} E.~A.,  2003, \mn@doi [\araa]
  {10.1146/annurev.astro.41.011802.094844}, \href
  {http://adsabs.harvard.edu/abs/2003ARA%26A..41...57L} {41, 57}

\bibitem[\protect\citeauthoryear{{Li}, {Liu}, {Hasegawa}  \& {Hirano}}{{Li}
  et~al.}{2017}]{Li17}
{Li} J.~I.,  {Liu} H.~B.,  {Hasegawa} Y.,   {Hirano} N.,  2017, \mn@doi [\apj]
  {10.3847/1538-4357/aa6f04}, \href
  {http://adsabs.harvard.edu/abs/2017ApJ...840...72L} {840, 72}

\bibitem[\protect\citeauthoryear{{Lin} et~al.,}{{Lin} et~al.}{2016}]{Lin16}
{Lin} Y.,  et~al., 2016, \mn@doi [\apj] {10.3847/0004-637X/828/1/32}, \href
  {http://adsabs.harvard.edu/abs/2016ApJ...828...32L} {828, 32}

\bibitem[\protect\citeauthoryear{{Liu}, {Quintana-Lacaci}, {Wang}, {Ho}, {Li},
  {Zhang}  \& {Zhang}}{{Liu} et~al.}{2012}]{Liu12}
{Liu} H.~B.,  {Quintana-Lacaci} G.,  {Wang} K.,  {Ho} P.~T.~P.,  {Li} Z.-Y.,
  {Zhang} Q.,   {Zhang} Z.-Y.,  2012, \mn@doi [\apj]
  {10.1088/0004-637X/745/1/61}, \href
  {http://adsabs.harvard.edu/abs/2012ApJ...745...61L} {745, 61}

\bibitem[\protect\citeauthoryear{{Liu}, {Galv{\'a}n-Madrid},
  {Jim{\'e}nez-Serra}, {Rom{\'a}n-Z{\'u}{\~n}iga}, {Zhang}, {Li}  \&
  {Chen}}{{Liu} et~al.}{2015}]{Liu15}
{Liu} H.~B.,  {Galv{\'a}n-Madrid} R.,  {Jim{\'e}nez-Serra} I.,
  {Rom{\'a}n-Z{\'u}{\~n}iga} C.,  {Zhang} Q.,  {Li} Z.,   {Chen} H.-R.,  2015,
  \mn@doi [\apj] {10.1088/0004-637X/804/1/37}, \href
  {http://adsabs.harvard.edu/abs/2015ApJ...804...37L} {804, 37}

\bibitem[\protect\citeauthoryear{{Maud}, {Hoare}, {Gibb}, {Shepherd}  \&
  {Indebetouw}}{{Maud} et~al.}{2013}]{Maud13b}
{Maud} L.~T.,  {Hoare} M.~G.,  {Gibb} A.~G.,  {Shepherd} D.,   {Indebetouw} R.,
   2013, \mn@doi [\mnras] {10.1093/mnras/sts049}, \href
  {http://adsabs.harvard.edu/abs/2013MNRAS.428..609M} {428, 609}

\bibitem[\protect\citeauthoryear{{Maud}, {Moore}, {Lumsden}, {Mottram},
  {Urquhart}  \& {Hoare}}{{Maud} et~al.}{2015}]{Maud+15}
{Maud} L.~T.,  {Moore} T.~J.~T.,  {Lumsden} S.~L.,  {Mottram} J.~C.,
  {Urquhart} J.~S.,   {Hoare} M.~G.,  2015, \mn@doi [\mnras]
  {10.1093/mnras/stv1635}, \href
  {http://adsabs.harvard.edu/abs/2015MNRAS.453..645M} {453, 645}

\bibitem[\protect\citeauthoryear{{Maud}, {Hoare}, {Galv{\'a}n-Madrid}, {Zhang},
  {de Wit}, {Keto}, {Johnston}  \& {Pineda}}{{Maud} et~al.}{2017}]{Maud+17}
{Maud} L.~T.,  {Hoare} M.~G.,  {Galv{\'a}n-Madrid} R.,  {Zhang} Q.,  {de Wit}
  W.~J.,  {Keto} E.,  {Johnston} K.~G.,   {Pineda} J.~E.,  2017, \mn@doi
  [\mnras] {10.1093/mnrasl/slx010}, \href
  {http://adsabs.harvard.edu/abs/2017MNRAS.467L.120M} {467, L120}

\bibitem[\protect\citeauthoryear{McKinney}{McKinney}{2010}]{Pandas}
McKinney W.,  2010, in van~der Walt S.,  Millman J.,  eds, Proceedings of the
  9th Python in Science Conference. pp 51 -- 56

\bibitem[\protect\citeauthoryear{{McMullin}, {Waters}, {Schiebel}, {Young}  \&
  {Golap}}{{McMullin} et~al.}{2007}]{McMullin07}
{McMullin} J.~P.,  {Waters} B.,  {Schiebel} D.,  {Young} W.,   {Golap} K.,
  2007, in {Shaw} R.~A.,  {Hill} F.,   {Bell} D.~J.,  eds,  Astronomical
  Society of the Pacific Conference Series Vol. 376, Astronomical Data Analysis
  Software and Systems XVI. p.~127

\bibitem[\protect\citeauthoryear{{Mendoza}, {Cant{\'o}}  \& {Raga}}{{Mendoza}
  et~al.}{2004}]{Mendoza04}
{Mendoza} S.,  {Cant{\'o}} J.,   {Raga} A.~C.,  2004, \rmxaa, \href
  {http://adsabs.harvard.edu/abs/2004RMxAA..40..147M} {40, 147}

\bibitem[\protect\citeauthoryear{{Motte}, {Bontemps}  \& {Louvet}}{{Motte}
  et~al.}{2017}]{Motte17}
{Motte} F.,  {Bontemps} S.,   {Louvet} F.,  2017, preprint, \href
  {http://adsabs.harvard.edu/abs/2017arXiv170600118M} {} (\mn@eprint {arXiv}
  {1706.00118})

\bibitem[\protect\citeauthoryear{{Nakamura} et~al.,}{{Nakamura}
  et~al.}{2012}]{Nakamura12}
{Nakamura} F.,  et~al., 2012, \mn@doi [\apj] {10.1088/0004-637X/746/1/25},
  \href {http://adsabs.harvard.edu/abs/2012ApJ...746...25N} {746, 25}

\bibitem[\protect\citeauthoryear{{Oey}}{{Oey}}{2011}]{Oey11}
{Oey} M.~S.,  2011, \mn@doi [\apjl] {10.1088/2041-8205/739/2/L46}, \href
  {http://adsabs.harvard.edu/abs/2011ApJ...739L..46O} {739, L46}

\bibitem[\protect\citeauthoryear{{Offner}, {Clark}, {Hennebelle}, {Bastian},
  {Bate}, {Hopkins}, {Moraux}  \& {Whitworth}}{{Offner}
  et~al.}{2014}]{Offner14}
{Offner} S.~S.~R.,  {Clark} P.~C.,  {Hennebelle} P.,  {Bastian} N.,  {Bate}
  M.~R.,  {Hopkins} P.~F.,  {Moraux} E.,   {Whitworth} A.~P.,  2014, \mn@doi
  [Protostars and Planets VI] {10.2458/azu_uapress_9780816531240-ch003}, \href
  {http://adsabs.harvard.edu/abs/2014prpl.conf...53O} {pp 53--75}

\bibitem[\protect\citeauthoryear{{Osorio}, {Anglada}, {Lizano}  \&
  {D'Alessio}}{{Osorio} et~al.}{2009}]{Osorio09}
{Osorio} M.,  {Anglada} G.,  {Lizano} S.,   {D'Alessio} P.,  2009, \mn@doi
  [\apj] {10.1088/0004-637X/694/1/29}, \href
  {http://adsabs.harvard.edu/abs/2009ApJ...694...29O} {694, 29}

\bibitem[\protect\citeauthoryear{{Patel} et~al.,}{{Patel}
  et~al.}{2005}]{Patel05}
{Patel} N.~A.,  et~al., 2005, \mn@doi [\nat] {10.1038/nature04011}, \href
  {http://adsabs.harvard.edu/abs/2005Natur.437..109P} {437, 109}

\bibitem[\protect\citeauthoryear{{Peretto} et~al.,}{{Peretto}
  et~al.}{2013}]{Peretto13}
{Peretto} N.,  et~al., 2013, \mn@doi [\aap] {10.1051/0004-6361/201321318},
  \href {http://adsabs.harvard.edu/abs/2013A%26A...555A.112P} {555, A112}

\bibitem[\protect\citeauthoryear{P\'erez \& Granger}{P\'erez \&
  Granger}{2007}]{IPython}
P\'erez F.,  Granger B.~E.,  2007, \mn@doi [Computing in Science and
  Engineering] {10.1109/MCSE.2007.53}, 9, 21

\bibitem[\protect\citeauthoryear{{P{\'e}rez} et~al.,}{{P{\'e}rez}
  et~al.}{2016}]{Perez16}
{P{\'e}rez} L.~M.,  et~al., 2016, \mn@doi [Science] {10.1126/science.aaf8296},
  \href {http://adsabs.harvard.edu/abs/2016Sci...353.1519P} {353, 1519}

\bibitem[\protect\citeauthoryear{{Peters}, {Banerjee}, {Klessen}, {Mac Low},
  {Galv{\'a}n-Madrid}  \& {Keto}}{{Peters} et~al.}{2010}]{Peters10}
{Peters} T.,  {Banerjee} R.,  {Klessen} R.~S.,  {Mac Low} M.-M.,
  {Galv{\'a}n-Madrid} R.,   {Keto} E.~R.,  2010, \mn@doi [\apj]
  {10.1088/0004-637X/711/2/1017}, \href
  {http://adsabs.harvard.edu/abs/2010ApJ...711.1017P} {711, 1017}

\bibitem[\protect\citeauthoryear{{Pringle}}{{Pringle}}{1981}]{Pringle81}
{Pringle} J.~E.,  1981, \mn@doi [\araa] {10.1146/annurev.aa.19.090181.001033},
  \href {http://adsabs.harvard.edu/abs/1981ARA%26A..19..137P} {19, 137}

\bibitem[\protect\citeauthoryear{{Purcell} et~al.,}{{Purcell}
  et~al.}{2006}]{Purcell05}
{Purcell} C.~R.,  et~al., 2006, \mn@doi [\mnras]
  {10.1111/j.1365-2966.2005.09921.x}, \href
  {http://adsabs.harvard.edu/abs/2006MNRAS.367..553P} {367, 553}

\bibitem[\protect\citeauthoryear{{Qu{\'e}nard}, {Bottinelli}  \&
  {Caux}}{{Qu{\'e}nard} et~al.}{2017}]{Quenard17}
{Qu{\'e}nard} D.,  {Bottinelli} S.,   {Caux} E.,  2017, \mn@doi [\mnras]
  {10.1093/mnras/stx404}, \href
  {http://adsabs.harvard.edu/abs/2017MNRAS.468..685Q} {468, 685}

\bibitem[\protect\citeauthoryear{{Remijan}, {Sutton}, {Snyder}, {Friedel},
  {Liu}  \& {Pei}}{{Remijan} et~al.}{2004}]{Remijan04}
{Remijan} A.,  {Sutton} E.~C.,  {Snyder} L.~E.,  {Friedel} D.~N.,  {Liu} S.-Y.,
    {Pei} C.-C.,  2004, \mn@doi [\apj] {10.1086/383120}, \href
  {http://adsabs.harvard.edu/abs/2004ApJ...606..917R} {606, 917}

\bibitem[\protect\citeauthoryear{{Rosen}, {Krumholz}, {McKee}  \&
  {Klein}}{{Rosen} et~al.}{2016}]{Rosen16}
{Rosen} A.~L.,  {Krumholz} M.~R.,  {McKee} C.~F.,   {Klein} R.~I.,  2016,
  \mn@doi [\mnras] {10.1093/mnras/stw2153}, \href
  {http://adsabs.harvard.edu/abs/2016MNRAS.463.2553R} {463, 2553}

\bibitem[\protect\citeauthoryear{{Salpeter}}{{Salpeter}}{1955}]{Salpeter55}
{Salpeter} E.~E.,  1955, \mn@doi [\apj] {10.1086/145971}, \href
  {http://adsabs.harvard.edu/abs/1955ApJ...121..161S} {121, 161}

\bibitem[\protect\citeauthoryear{{Sana}}{{Sana}}{2017}]{Sana17}
{Sana} H.,  2017, preprint, \href
  {http://adsabs.harvard.edu/abs/2017arXiv170301608S} {} (\mn@eprint {arXiv}
  {1703.01608})

\bibitem[\protect\citeauthoryear{{S{\'a}nchez-Monge}
  et~al.,}{{S{\'a}nchez-Monge} et~al.}{2013}]{SanchezMonge13}
{S{\'a}nchez-Monge} {\'A}.,  et~al., 2013, \mn@doi [\aap]
  {10.1051/0004-6361/201321134}, \href
  {http://adsabs.harvard.edu/abs/2013A%26A...552L..10S} {552, L10}

\bibitem[\protect\citeauthoryear{{Schmiedeke} et~al.,}{{Schmiedeke}
  et~al.}{2016}]{Schmiedeke16}
{Schmiedeke} A.,  et~al., 2016, \mn@doi [\aap] {10.1051/0004-6361/201527311},
  \href {http://adsabs.harvard.edu/abs/2016A%26A...588A.143S} {588, A143}

\bibitem[\protect\citeauthoryear{{Schneider}, {Csengeri}, {Bontemps}, {Motte},
  {Simon}, {Hennebelle}, {Federrath}  \& {Klessen}}{{Schneider}
  et~al.}{2010}]{Schneider10}
{Schneider} N.,  {Csengeri} T.,  {Bontemps} S.,  {Motte} F.,  {Simon} R.,
  {Hennebelle} P.,  {Federrath} C.,   {Klessen} R.,  2010, \mn@doi [\aap]
  {10.1051/0004-6361/201014481}, \href
  {http://adsabs.harvard.edu/abs/2010A%26A...520A..49S} {520, A49}

\bibitem[\protect\citeauthoryear{{Sch{\"o}ier}, {van der Tak}, {van Dishoeck}
  \& {Black}}{{Sch{\"o}ier} et~al.}{2005}]{Schoier+05}
{Sch{\"o}ier} F.~L.,  {van der Tak} F.~F.~S.,  {van Dishoeck} E.~F.,   {Black}
  J.~H.,  2005, \mn@doi [\aap] {10.1051/0004-6361:20041729}, \href
  {http://esoads.eso.org/abs/2005A%26A...432..369S} {432, 369}

\bibitem[\protect\citeauthoryear{{Smith}, {Longmore}  \& {Bonnell}}{{Smith}
  et~al.}{2009}]{Smith09}
{Smith} R.~J.,  {Longmore} S.,   {Bonnell} I.,  2009, \mn@doi [\mnras]
  {10.1111/j.1365-2966.2009.15621.x}, \href
  {http://adsabs.harvard.edu/abs/2009MNRAS.400.1775S} {400, 1775}

\bibitem[\protect\citeauthoryear{{Stier} et~al.,}{{Stier}
  et~al.}{1984}]{Stier84}
{Stier} M.~T.,  et~al., 1984, \mn@doi [\apj] {10.1086/162342}, \href
  {http://adsabs.harvard.edu/abs/1984ApJ...283..573S} {283, 573}

\bibitem[\protect\citeauthoryear{{Ulrich}}{{Ulrich}}{1976}]{Ulrich76}
{Ulrich} R.~K.,  1976, \mn@doi [\apj] {10.1086/154840}, \href
  {http://adsabs.harvard.edu/abs/1976ApJ...210..377U} {210, 377}

\bibitem[\protect\citeauthoryear{{Van Der Walt}, {Colbert}  \&
  {Varoquaux}}{{Van Der Walt} et~al.}{2011}]{Numpy}
{Van Der Walt} S.,  {Colbert} S.~C.,   {Varoquaux} G.,  2011, preprint, \href
  {http://adsabs.harvard.edu/abs/2011arXiv1102.1523V} {} (\mn@eprint {arXiv}
  {1102.1523})

\bibitem[\protect\citeauthoryear{{V{\'a}zquez-Semadeni},
  {Gonz{\'a}lez-Samaniego}  \& {Col{\'{\i}}n}}{{V{\'a}zquez-Semadeni}
  et~al.}{2017}]{VS17}
{V{\'a}zquez-Semadeni} E.,  {Gonz{\'a}lez-Samaniego} A.,   {Col{\'{\i}}n} P.,
  2017, \mn@doi [\mnras] {10.1093/mnras/stw3229}, \href
  {http://adsabs.harvard.edu/abs/2017MNRAS.467.1313V} {467, 1313}

\bibitem[\protect\citeauthoryear{{Vincke}, {Breslau}  \& {Pfalzner}}{{Vincke}
  et~al.}{2015}]{Vincke15}
{Vincke} K.,  {Breslau} A.,   {Pfalzner} S.,  2015, \mn@doi [\aap]
  {10.1051/0004-6361/201425552}, \href
  {http://adsabs.harvard.edu/abs/2015A%26A...577A.115V} {577, A115}

\bibitem[\protect\citeauthoryear{{Vorobyov}, {Zakhozhay}  \&
  {Dunham}}{{Vorobyov} et~al.}{2013}]{Vorobyov13}
{Vorobyov} E.~I.,  {Zakhozhay} O.~V.,   {Dunham} M.~M.,  2013, \mn@doi [\mnras]
  {10.1093/mnras/stt970}, \href
  {http://adsabs.harvard.edu/abs/2013MNRAS.433.3256V} {433, 3256}

\bibitem[\protect\citeauthoryear{{Whitney}, {Wood}, {Bjorkman}  \&
  {Wolff}}{{Whitney} et~al.}{2003}]{Whitney03}
{Whitney} B.~A.,  {Wood} K.,  {Bjorkman} J.~E.,   {Wolff} M.~J.,  2003, \mn@doi
  [\apj] {10.1086/375415}, \href
  {http://adsabs.harvard.edu/abs/2003ApJ...591.1049W} {591, 1049}

\bibitem[\protect\citeauthoryear{{Wilner}, {Wright}  \& {Plambeck}}{{Wilner}
  et~al.}{1994}]{Wilner94}
{Wilner} D.~J.,  {Wright} M.~C.~H.,   {Plambeck} R.~L.,  1994, \mn@doi [\apj]
  {10.1086/173757}, \href {http://adsabs.harvard.edu/abs/1994ApJ...422..642W}
  {422, 642}

\bibitem[\protect\citeauthoryear{{Wright}, {Hull}, {Pillai}, {Zhao}  \&
  {Sandell}}{{Wright} et~al.}{2014}]{Wright14}
{Wright} M.~C.~H.,  {Hull} C.~L.~H.,  {Pillai} T.,  {Zhao} J.-H.,   {Sandell}
  G.,  2014, \mn@doi [\apj] {10.1088/0004-637X/796/2/112}, \href
  {http://adsabs.harvard.edu/abs/2014ApJ...796..112W} {796, 112}

\bibitem[\protect\citeauthoryear{{Yen} et~al.,}{{Yen} et~al.}{2017}]{Yen17}
{Yen} H.-W.,  et~al., 2017, preprint, \href
  {http://adsabs.harvard.edu/abs/2017arXiv170802384Y} {} (\mn@eprint {arXiv}
  {1708.02384})

\bibitem[\protect\citeauthoryear{{Zhang}, {Hunter}, {Sridharan}  \&
  {Ho}}{{Zhang} et~al.}{2002}]{Zhang02}
{Zhang} Q.,  {Hunter} T.~R.,  {Sridharan} T.~K.,   {Ho} P.~T.~P.,  2002,
  \mn@doi [\apj] {10.1086/338278}, \href
  {http://adsabs.harvard.edu/abs/2002ApJ...566..982Z} {566, 982}

\bibitem[\protect\citeauthoryear{{Zinnecker} \& {Yorke}}{{Zinnecker} \&
  {Yorke}}{2007}]{ZY07}
{Zinnecker} H.,  {Yorke} H.~W.,  2007, \mn@doi [\araa]
  {10.1146/annurev.astro.44.051905.092549}, \href
  {http://adsabs.harvard.edu/abs/2007ARA%26A..45..481Z} {45, 481}

\bibitem[\protect\citeauthoryear{{de Wit}, {Hoare}, {Oudmaijer}  \&
  {Mottram}}{{de Wit} et~al.}{2007}]{deWit07}
{de Wit} W.~J.,  {Hoare} M.~G.,  {Oudmaijer} R.~D.,   {Mottram} J.~C.,  2007,
  \mn@doi [\apjl] {10.1086/525253}, \href
  {http://adsabs.harvard.edu/abs/2007ApJ...671L.169D} {671, L169}

\bibitem[\protect\citeauthoryear{{de Wit}, {Hoare}, {Oudmaijer}  \&
  {Lumsden}}{{de Wit} et~al.}{2010}]{deWit10}
{de Wit} W.~J.,  {Hoare} M.~G.,  {Oudmaijer} R.~D.,   {Lumsden} S.~L.,  2010,
  \mn@doi [\aap] {10.1051/0004-6361/200913209}, \href
  {http://adsabs.harvard.edu/abs/2010A%26A...515A..45D} {515, A45}

\bibitem[\protect\citeauthoryear{{van der Tak} \& {Menten}}{{van der Tak} \&
  {Menten}}{2005}]{vdTMenten05}
{van der Tak} F.~F.~S.,  {Menten} K.~M.,  2005, \mn@doi [\aap]
  {10.1051/0004-6361:20052872}, \href
  {http://adsabs.harvard.edu/abs/2005A%26A...437..947V} {437, 947}

\bibitem[\protect\citeauthoryear{{van der Tak}, {van Dishoeck}, {Evans}  \&
  {Blake}}{{van der Tak} et~al.}{2000}]{vdT00}
{van der Tak} F.~F.~S.,  {van Dishoeck} E.~F.,  {Evans} II N.~J.,   {Blake}
  G.~A.,  2000, \mn@doi [\apj] {10.1086/309011}, \href
  {http://adsabs.harvard.edu/abs/2000ApJ...537..283V} {537, 283}

\makeatother
\end{thebibliography}

\appendix

\section{Model and Observed Channel Maps}\label{ap:chanmaps}

\begin{figure*}
 \centering
 \includegraphics[width = 1\textwidth, height = 0.93\textheight]{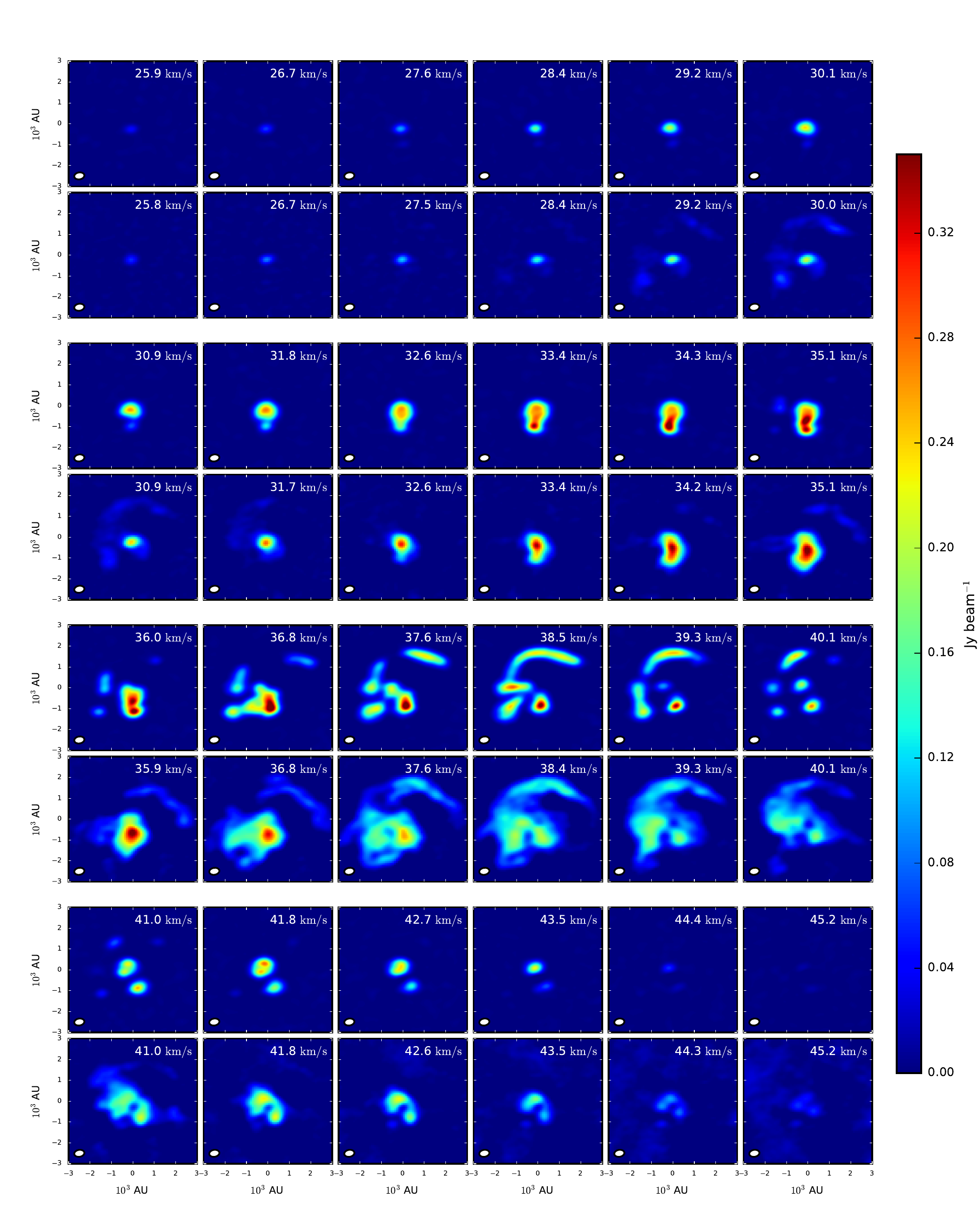} 
 \caption{Channel maps for CH$_3$CN $J=19-18$, $K=4$ in the model (odd rows) and the observations (even rows). The velocity range is from 25.9 km s$^{-1}$ to 45.2 km s$^{-1}$, with a step of 0.84 km s$^{-1}$ (the separation between channels in the model and the data is 0.42 km s$^{-1}$). The beam size is shown in the lower left corner of the panels. A color bar in the right side of the figure indicates the corresponding intensity.}  
\label{fig:channels}
\end{figure*}

\section{Model Library}\label{ap:models}

To carry out the modelling of W33A MM1 we developed a set of libraries to create analytical distributions of physical properties such as density, temperature, velocity, and molecular abundance. 
The package is modular, intended to be user-friendly, and entirely written in Python. The basic branch of the package allows to reproduce a single (star-forming) region based on a group of standard models as the ones referred to in Section \ref{sec:physmodel}, as well as simpler analytical distributions such as power-law and homogeneous profiles. The user has the possibility of defining some combinations of models without the need of defining a more complex `global' grid, for example, a Pringle disc embedded in an Ulrich envelope or in a user-defined power law, or defining gaps and cavities within a region of the previously invoked model.  

For advanced setups, the package allows the user to model sets of individual sources together within a global grid, as it was done in this paper. The tools for these advanced feature include libraries that are able to define the overlapping process of individual models, as well as their translations and rotations. The global grid that allocates all the user-defined regions is built on the go. 
Filamentary structures with cylindrical or parabolic shapes can also be generated within the global grid. 

The output of the modelling package includes data tables with numerical values of density, temperature, velocity, abundance, and gas-to-dust ratio for each individual model, and a global data table with overlapped physical properties (as explained in Section \ref{sec:GlobalGrid}) in the case that the user decides to join two or more individual models.

The output was adapted to be the input of the Line Modelling engine software \citep[{\small LIME},][]{Brinch+10} to obtain predictions of the line and continuum radiation observed from the model. A header file that reads the input physical properties adequately for LIME is included in the package. An illustrative example of the use of the package, from the model definitions to its integration with LIME, is also included. 

The package and its documentation are available through GitHub (\href{https://github.com/andizq/star-forming-regions}{https://github.com/andizq/star-forming-regions}). We foresee to update the package with more features in the near future, including model `ingestors' for more recent versions of LIME and other radiative transfer codes, as well as free-free and recombination line calculations. We kindly ask the reader to refer to this work or to the related publications of future developments if this software has been useful for their research. 

\end{document}